\documentclass[twocolumn]{aastex62}

\def\teff{\mbox{T$_{\rm eff}$}}
\def\logg{\mbox{log~{\it g}}}
\def\vmicro{\mbox{$\xi_{\rm turb}$}}
\def\kmsec{\mbox{km~s$^{\rm -1}$}}

\usepackage{epstopdf}
\usepackage{amsmath}

\accepted{ }

\submitjournal{ApJ}

\shorttitle{Chemical Compositions of RR Lyrae Stars in NGC~3201}
\shortauthors{Magurno et al.}

\begin{document}

\title{Chemical Compositions of Field and Globular Cluster RR~Lyrae Stars: I. NGC~3201\footnote{This paper includes data gathered with the 6.5 meter Magellan 
Telescopes located at Las Campanas Observatory, Chile.}}

\correspondingauthor{Davide Magurno}
\email{davide.magurno@roma2.infn.it}

\author{D. Magurno}
\affiliation{University of Roma Tor Vergata, via della Ricerca Scientifica 1, 00133, Roma, Italy}
\affiliation{INAF Osservatorio Astronomico di Roma, via Frascati 33, 00040, Monte Porzio Catone RM, Italy}

\author{C. Sneden}
\affiliation{Department of Astronomy and McDonald Observatory, The University of Texas, Austin, TX 78712, USA}

\author{V. F. Braga}
\affiliation{Instituto Milenio de Astrof{\'i}sica, Santiago, Chile}
\affiliation{Departamento de F{\'i}sica, Facultad de Ciencias Exactas, Universidad Andr{\'e}s Bello, Fern{\'a}ndez Concha 700, Las Condes, Santiago, Chile}

\author{G. Bono}
\affiliation{University of Roma Tor Vergata, via della Ricerca Scientifica 1, 00133, Roma, Italy}
\affiliation{INAF Osservatorio Astronomico di Roma, via Frascati 33, 00040, Monte Porzio Catone RM, Italy}

\author{M. Mateo}
\affiliation{Department of Astronomy, University of Michigan, 1085 S. University, Ann Arbor, MI 48109, USA}

\author{S. E. Persson}
\affiliation{Observatories of the Carnegie Institution for Science, 813 Santa Barbara Street, Pasadena, CA 91101, USA}

\author{M. Dall'Ora}
\affiliation{INAF Osservatorio Astronomico di Capodimonte, Salita Moiariello 16, 80131 Napoli, Italy}

\author{M. Marengo}
\affiliation{Department of Physics and Astronomy, Iowa State University, A313E Zaffarano, Ames, IA 50010, USA }

\author{M. Monelli}
\affiliation{IAC - Instituto de Astrofisica de Canarias, Calle Via Lactea, E38200 La Laguna, Tenerife, Espana}

\author{J. R. Neeley}
\affiliation{Department of Physics, Florida Atlantic University, Boca Raton, FL 33431, USA}

\begin{abstract}
We present a detailed spectroscopic analysis of horizontal branch stars in the 
globular cluster NGC~3201. We collected 
optical (4580--5330~\AA), high resolution ($\sim$34,000), high signal-to-noise ratio ($\sim$200) 
spectra for eleven RR~Lyrae stars and one red horizontal branch star with the 
multifiber spectrograph M2FS at the 6.5m Magellan telescope at the Las Campanas Observatory. 
From measured equivalent widths we derived 
atmospheric parameters and abundance ratios for $\alpha$ (Mg, Ca, Ti), iron peak (Sc, Cr, Ni, Zn) and 
s-process (Y) elements. We found that NGC~3201 is a homogeneous, mono-metallic
([Fe/H]=$-1.47\pm0.04$), $\alpha$-enhanced ([$\alpha$/Fe]=$0.37\pm0.04$) cluster. 
The relative abundances of the iron peak and s-process elements were found to be consistent with solar values.
In comparison with other large stellar samples, NGC~3201 RR~Lyraes have 
similar chemical enrichment histories as do those of other old 
(t$\ge$10 Gyr) Halo components (globular clusters, 
red giants, blue and red horizontal branch stars, RR Lyraes).
We also provided a new average radial velocity estimate for NGC~3201 by using 
a template velocity curve to overcome the limit of single epoch measurements 
of variable stars:  V$_{\text{rad}}=494\pm2$ km~s$^{-1}$ 
($\sigma$=8 km~s$^{-1}$).

\end{abstract}


\section{Introduction} \label{sec:intro}

Dating back to \cite{baade58a}, RR Lyrae stars (RRL) have 
played a fundamental role as tracers of old (t$>$10 Gyr) stellar populations. 
RRLs are ubiquitous, having been identified both in gas-poor and 
in gas-rich stellar systems.
Moreover, they can be easily identified thanks to a particular coupling 
between pulsation period and shape/amplitude of their optical light curves.   
Therefore they have been extensively used
to investigate 
the early formation and the spatial structure of the Galactic 
Bulge \citep{pietrukowicz15} and of the Galactic
Halo \citep{drake13a,torrealba15}.
The RRLs in globulars have been widely used not only to constrain the 
evolutionary properties of old, low-mass, central helium burning stars, 
but also to investigate the impact that the intrinsic parameters (metallicity) 
and the environment have on the topology of the instability strip and on   
their pulsation properties \citep{oosterhoff39,vanalbada73,caputo97,bono07}. 

RRLs are also very good distance indicators. 
Dating back once again to \cite{baade55} and to \cite{sandage58} it was
shown that RRLs have a well defined visual magnitude--metallicity relation.  
Their use as standard candles became even more compelling thanks to the 
empirical discovery by \cite{longmore86,longmore90} that RRLs have
near-infrared Period-Luminosity correlations. 
More recent empirical and theoretical evidence indicate that in the near infrared (NIR) 
they obey Period--Luminosity--Metallicity (PLZ) 
relations \citep{bono03c,marconi15,braga15,neeley17,braga18}.

The RRLs have also played a crucial role in the investigation of the 
spatial distribution of old stellar populations in nearby dwarf galaxies 
(Magellanic Clouds: \citealt{soszynski09a}; Carina: \citealt{coppola13}; 
Sculptor: \citealt{martinezvazquez16b}). 
HST optical photometry played a fundamental role in 
detecting and tracing RRLs in satellites of M31 
\citep{clementini01,pritzl02,monelli17}, and in galaxies of the Sculptor 
group \citep{dacosta10}.
The pulsation properties of RRLs in globulars, and in Local Group (d$<$1 Mpc) 
and Local Volume (d$<$10 Mpc) galaxies, can be adopted to constrain the early 
formation and evolution of the Galactic spheroid 
\citep{stetson14a,fiorentino15a}.

RRLs in globular clusters are especially useful for 
several reasons.  
First, the ages and the chemical compositions of many clusters are well known. 
In particular, the iron metallicities, $\alpha$ and neutron capture elements 
have been studied extensively, \citep[e.g,][]{carretta09}. 
Second, the evolutionary status and the topology of the instability
strip is also well established \citep{walker17}.
The globulars hosting a sizable sample of RRLs allow us to investigate
the regions of the instability strip in which variables pulsate either
as first overtones (hotter) or as fundamentals (cooler). 
Moreover, we can also estimate the width in temperature of the region 
in which RRLs pulsate simultaneously in the first overtone and in the
fundamental mode, i.e., the so-called mixed mode pulsators. 
Finally, evolved cluster RRLs can be more easily identified, since they 
attain luminosities that are systematically brighter than the zero-age
horizontal branch (ZAHB) luminosity level.
 
Metallicities and detailed abundance ratios of individual 
stars are crucial not only to provide more accurate individual distance 
determinations, but also to trace the early chemical enrichment of old stellar 
populations \citep{monelli12b,martinezvazquez16a}. 
Metallicities of field RRLs have been derived from the large and homogeneous SDSS DR8 sample of medium 
resolution spectra \citep{lee11,drake13a}, using several different techniques,
mostly based on photometric indices \citep{mateu12} or on metallicity 
indicators like the Ca II K lines ($\Delta S$, \citealt{preston59,layden94}). 
Recently, metallicities for field RRLs have been estimated using several 
spectroscopic indicators and collected at different pulsation phases.

High resolution spectroscopic analyses of field RRLs are 
currently limited to $\sim$140 stars \citep[e.g.,][]{clementini95,liu13,pancino15,chadid17,sneden17,andrievsky18}.
These studies have lagged compared with those of other groups of variables 
stars (Classical Cepheids, Miras) for many reasons:

a)
Pulsation periods of RRLs range from a few hours for 
first-overtone, RRc, pulsators to almost one day for fundamental, RRab, pulsators.
This means that the exposure time to collect spectra can hardly be longer 
than 30--45 minutes to avoid velocity smearing of spectra. 
To acquire high resolution and S/N spectra typically requires using 4--8 m class 
telescopes, or co-adding spectra obtained over many pulsation cycles by 
smaller telescopes.
This requires well-known pulsational timing to avoid overlap of 
different phases, and in turn different physical properties.

b)
RRab stars experience several non-linear phenomena 
during their pulsation cycles.
The formation and propagation of strong shocks across the rising-light 
branch cause line doubling and P Cygni profiles \citep{preston59,preston64}. 
During these pulsation phases especially the assumption of quasi-static 
atmospheres is no longer valid, since the line formation takes place in a 
medium affected by sharp temperature and density gradients \citep{bono94c}.

c)
Up to 50\% of RRab stars exhibit a Blazhko effect
\citep{kolenberg10b,benko14}, i.e. a quasi-periodic (tens to hundreds of days) 
modulation of the lightcurve amplitude and period 
\citep{jurcsik09,kolenberg10}.
Many hypothesis have been formulated to explain the Blazhko effect. Recently, \cite{buchler11} suggested that the modulation is the consequence of 
resonance between the fundamental and the ninth overtone pulsation modes,
however, we still lack agreement on a convincing physical explanation.

d)
The RRLs cover a very broad metallicity range. 
Current estimates suggest a range from [Fe/H]~$\simeq -2.9$ \citep{govea14} 
to [Fe/H]~$\simeq$~0.1 \citep{chadid17}\footnote{We adopted the standard notation,
[X/H]=A(X)-A$_\sun$(X), where A(X)=log(N$_X$)$-12$. Solar abundances
refer to \cite{asplund09} within the text.}.
The identification and measurement of individual atomic lines requires 
high spectral resolution. 
Such lines are not plentiful in RRLs, which have temperatures 
\teff~$\simeq$6000$-$8000~K and surface gravities \logg~$\simeq$~2.5$\pm$0.5 
\citep{bono94b,marconi15}.

In this paper we report an abundance analysis on eleven RRLs 
and one red horizontal branch (RHB) star in the globular cluster NGC~3201, 
using high resolution optical spectra collected with M2FS at Magellan.
We have derived [Fe/H] metallicities and abundance ratios [X/Fe] of 
$\alpha$-elements (Mg, Ca, Ti), iron peak elements (Sc, Cr, Ni, Zn)
and one s-process element (Y).
This GC has been widely investigated using giant branch (RGB, AGB) stars 
\citep[e.g.,][]{carretta09} and it is generally accepted to be a mono-metallic 
cluster ([Fe/H] $\sim -1.5$), 
but an extended analysis of its RRLs has not been 
previously done.
The only RRL-based analysis of NGC~3201 was performed by \cite{smith83},
who estimated iron abundances using the $\Delta S$ technique.
Therefore, our main goal is to obtain a new independent abundance analysis for 
NGC~3201 based on RRLs and to compare it with results available in the 
literature for RGB and AGB stars.
In Section~\ref{sec:instr} we describe the instrument and
the data sample.
In Section~\ref{sec:veloc} we discuss the radial velocity
analysis.
Section~\ref{sec:abund} focuses on abundance determinations and comparison 
with other globular clusters and field stars.
Finally, Section~\ref{sec:fine} summarizes the current results. 

\begin{deluxetable*}{ccccccclccc}
\tablecaption{Photometric parameters for the sample stars in NGC~3201. \label{tab:position}}
\tablewidth{0pt}
\tablehead{
\colhead{ID} &
\colhead{$\alpha$} &
\colhead{$\delta$} &
\colhead{p (days)} &
\colhead{HJD\tablenotemark{a}} &
\colhead{HJD$_0$\tablenotemark{a}} & 
\colhead{phase} &
\colhead{type} &
\colhead{$\langle$V$\rangle$} &
\colhead{Vamp} &
\colhead{Reference}
}
\startdata
V3  &  10:17:54.47 & -46:25:25.4  &  0.59939921 & 7079.77870   &    4123.15325    &  0.65   &  RRab                   &  14.90  &  0.67  & N,L03 \\
V6  &  10:17:26.09 & -46:27:02.3  &  0.52561240 & 7079.77870   &    6040.73538    &  0.82   &  RRab                   &  14.74  &  0.93  & N,L03 \\
V14 &  10:17:22.42 & -46:22:31.5  &  0.50929203 & 7079.77870   &    6718.64605    &  0.09   &  RRab                   &  14.95  &  1.12  & N \\
V26 &  10:17:58.09 & -46:27:01.8  &  0.56896113 & 7079.75386   &    6718.77606    &  0.45   &  RRab                   &  14.90  &  0.92  & N \\
V37 &  10:17:30.69 & -46:25:55.7  &  0.57699328 & 7079.75386   &    4123.08217    &  0.27   &  RRab                   &  14.78  &  0.78  & N \\
V38 &  10:17:31.42 & -46:25:41.1  &  0.50909990 & 7079.77870   &    6718.57488    &  0.49   &  RRab\tablenotemark{b}  &  14.76  &  1.04  & N \\
V41 &  10:18:05.02 & -46:24:14.8  &  0.66532664 & 7079.75386   &    7961.19997    &  0.17   &  RRab                   &  14.73  &  0.40  & ASN,N \\
V47 &  10:17:47.56 & -46:20:41.7  &  0.52086843 & 7079.75386   &    7960.97248    &  0.17   &  RRab                   &  14.60  &  0.89  & ASN,L03 \\
V57 &  10:18:04.87 & -46:25:54.6  &  0.59343497 & 7079.77870   &    6718.65687    &  0.53   &  RRab                   &  14.83  &  0.74  & N \\
V73 &  10:17:25.20 & -46:23:15.2  &  0.51995506 & 7079.75386   &    6718.84703    &  0.11   &  RRab                   &  14.75  &  1.22  & N \\
V83 &  10:17:54.76 & -46:21:54.4  &  0.54520516 & 7079.77870   &    6718.79059    &  0.11   &  RRab                   &  14.79  &  1.23  & N \\
94180 &   10:17:46.54 & -46:27:17.3  &  \ldots   & 7079.75386   &   \ldots    & \ldots    &     RHB               &  13.86  &   \ldots  & N \\
\enddata
\tablenotetext{}{References: N=Neeley et al., in preparation; L03=\cite{layden03}; ASN=ASAS-SN \citep{shappee14,jayasinghe18}}
\tablenotetext{a}{2450000+}
\tablenotetext{b}{Blazhko}
\end{deluxetable*}

\begin{figure}
\plotone{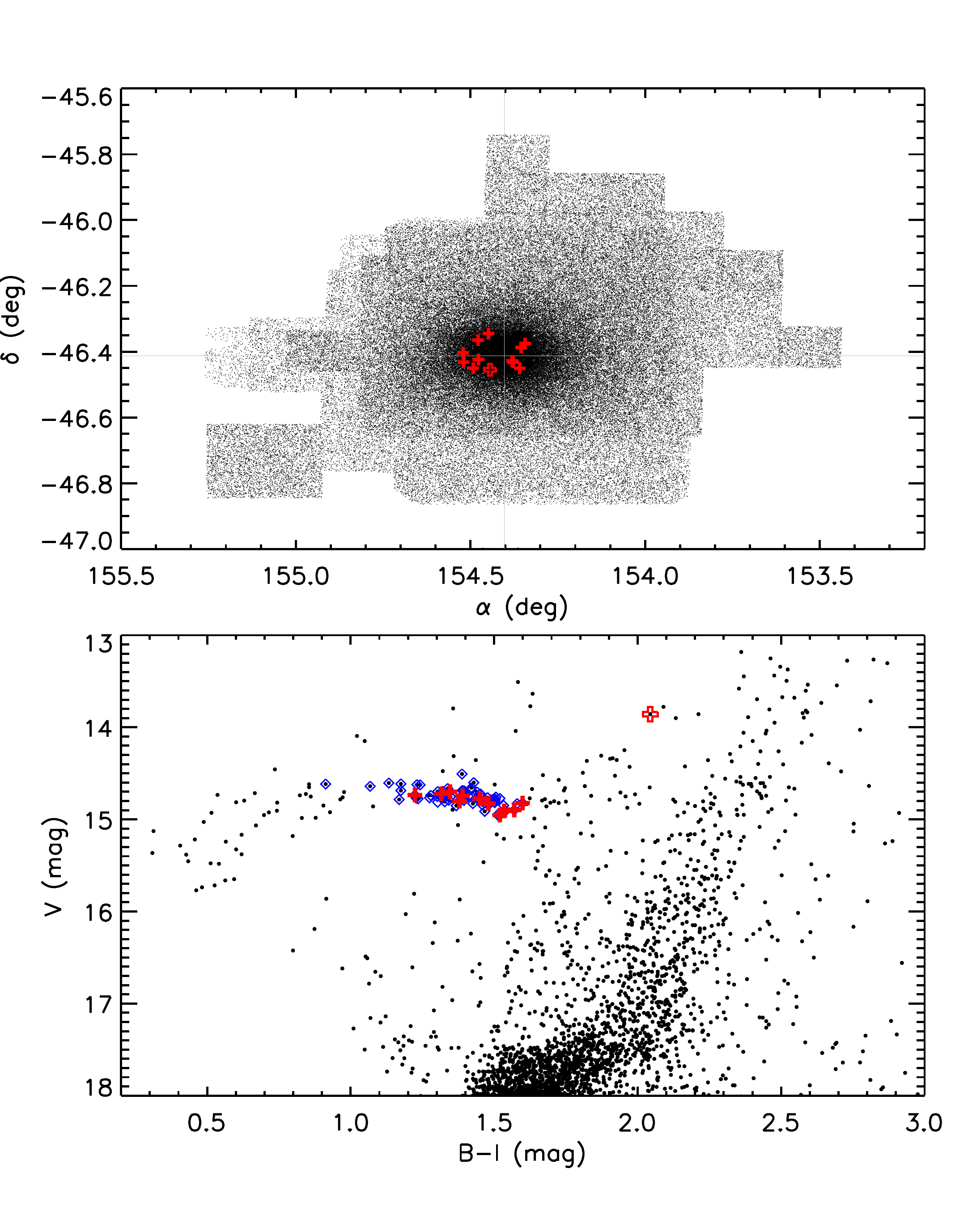}
\caption{Top panel-- Radial distribution of the 12 spectroscopic stars for which we collected high-resolution optical spectra (RRLS: filled red crosses; RHB: empty red cross) in the globular cluster NGC~3201 (black dots, photometry from Neeley et al., in preparation). Bottom panel-- V,B$-$I CMD of NGC~3201. Known cluster RRLs are shown with blue diamonds. \label{fig:CMD}}
\end{figure}

\begin{figure}
\plotone{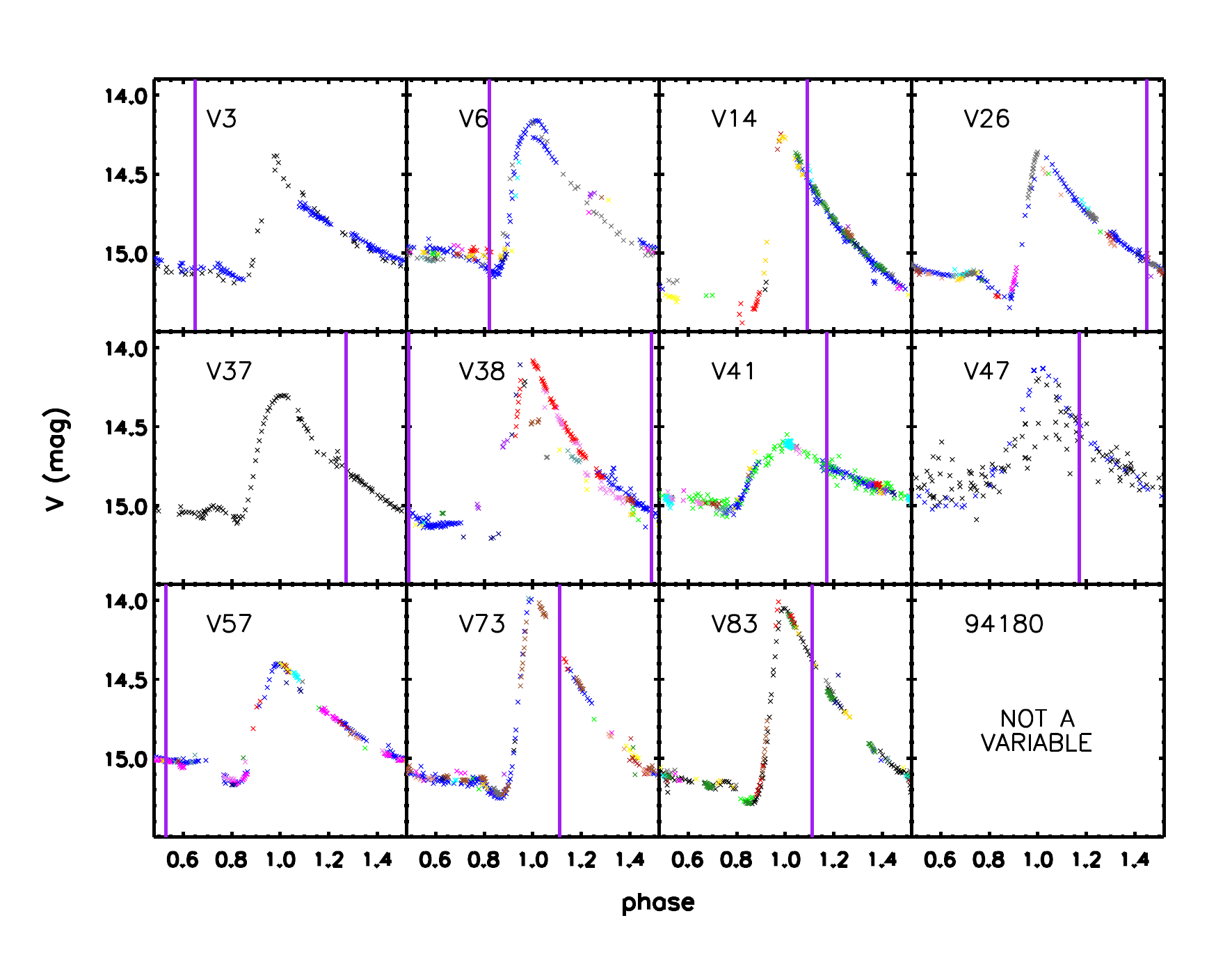}
\caption{V band lightcurves for the 11 sample RRLs. Color coding highlights different datasets. Purple vertical lines show the observed phases for the individual spectra. \label{fig:curve}}
\end{figure}

\section{Instrument and data sample} \label{sec:instr}
In February 2015,  we collected spectra for eleven RRLs and one RHB star in the globular cluster
NGC~3201 (Figure~\ref{fig:CMD})
using the Michigan/Magellan Fiber System (M2FS, \citealt{mateo12}) installed
at the Magellan/Clay 6.5m telescope at the Las Campanas Observatory in Chile.
The spectrograph configuration used an order-isolation filter to limit the
spectral coverage to 4580--5330~\AA\ in eleven overlapping echelle orders.
A total of eight stars could be observed with one setup with each of the two
camera/detector units, or 16 stars in all.
The spectrograph entrance slit size was 95~$\micron$, which yielded spectra
with resolving power $R$~$\equiv$ $\lambda/\Delta\lambda$ $\simeq$~34,000.

All the observed variables are RRab, including V38 which is 
also a Blazkho candidate (\citealt{layden03}, Neeley et al. in preparation).
It is possible to deduce the nature of V38 from its characteristic lightcurve 
(Figure~\ref{fig:curve}).
Unfortunately, only for two stars (V41 and V47) is very recent photometry available 
by ASAS-SN \citep{shappee14,jayasinghe18}. 
The other variables have photometry from one to more than ten years older than 
our spectra, so that the phase determination could be affected by some inaccuracy.
Average magnitudes and new estimated periods are listed in
Table~\ref{tab:position}.

\section{Radial Velocities} \label{sec:veloc}
\begin{figure}
\plotone{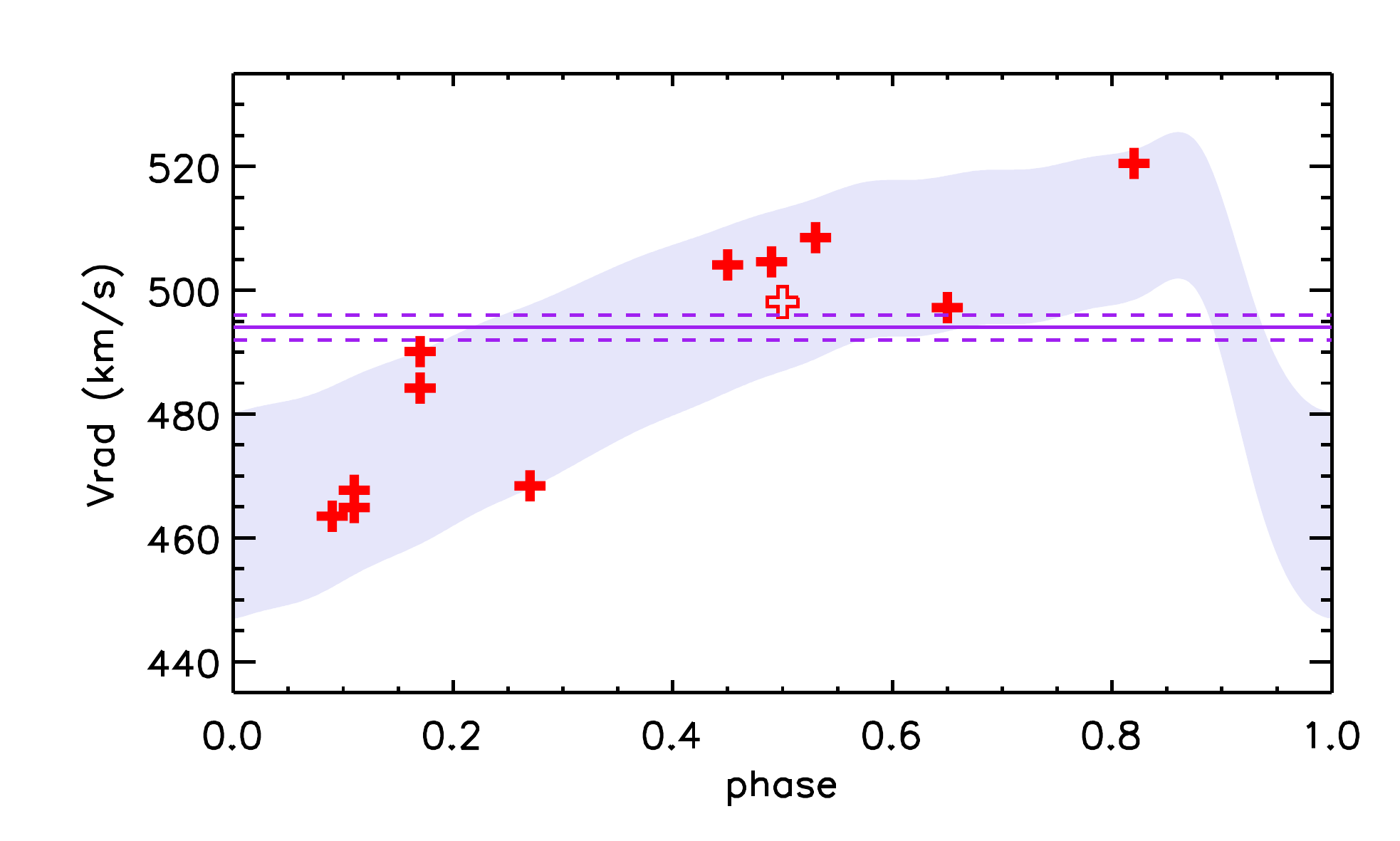}
\caption{Measured instantaneous radial velocity for the sample stars (red crosses). The empty cross marks the RHB star at the arbitrary phase 0.5. The shaded area shows an ensemble of the velocity curves obtained with the template by \cite{sesar12}. The average cluster velocity and errors on the mean are shown with purple lines. \label{fig:vrad}}
\end{figure}

The radial velocities of the sample were estimated using 
the task {\it fxcor} in IRAF \citep{tody86,tody93}\footnote{
IRAF is distributed by the National Optical Astronomy Observatories, which 
are operated by the Association of Universities for Research in Astronomy, 
Inc., under cooperative agreement with the National Science Foundation.},
by cross-correlation of the overall spectral range with a synthetic spectrum.
The synthetic spectrum was generated with the MOOG
driver {\it synth} \citep{sneden73}, with an \textquotedblleft average
RRL\textquotedblright\ parameters setting (\teff=6500~K, \logg=2.5,
\vmicro=3.0~km~s$^{-1}$), and a metallicity comparable with the
literature estimates for NGC~3201 ([Fe/H]=$-$1.5).  This computed spectrum
was then smoothed to the M2FS resolution (R=34,000).
%An accurate NGC~3201 cluster radial velocity was determined 
%by \cite{cote95}: 494.0$\pm$0.2 km~s$^{-1}$, using 399 non-variable stars.
An accurate NGC~3201 cluster radial velocity has been recently determined 
by \cite{ferraro18}: 494.5$\pm$0.4 km~s$^{-1}$, using 454 non-variable stars.
This very large value makes it a good indicator of cluster membership for
individual stars.
However, we are dealing with variable stars and so the instantaneous velocities 
(V$_{\text{rad}}$, Table~\ref{tab:obs_param})
are not good representative of the average cluster velocity. 
The mean velocity obtained from our single epoch measurements 
is 489$\pm$6 km~s$^{-1}$.
To correct for the pulsational velocities we reconstructed the velocity curve
for each star over the entire pulsation cycle by using a template \citep{sesar12}. 
The first step was
to select the template radial velocity among those available, 
which 
are based on hydrogen (H$_{\alpha}$, H$_{\beta}$, H$_{\gamma}$) or metallic lines. 
We chose the latter of these since our radial velocities are 
based on metallic lines. 
After that, we used equation 5 in \citet{sesar12} to 
rescale the template to the appropriate amplitude 
of each star (Table~\ref{tab:position}, Figure~\ref{fig:curve}).
Using the epoch of maximum light displayed in Table~\ref{tab:obs_param}, 
we have anchored the template radial velocity to the measured point and derived 
the systemic velocity as the integral average velocity along the pulsation cycle 
(V$_{\gamma}$, Table~\ref{tab:obs_param}).
Finally, we estimated the cluster radial velocity as 494$\pm$2 km~s$^{-1}$, independent of phase,
with a standard deviation of 8 km~s$^{-1}$.
However, assumptions about the reconstruction of the velocity curves
and errors in the estimate of the epoch of maximum light affect the template results.
Figure~\ref{fig:vrad} shows the instantaneous radial velocities for the 12 stars 
in our sample and an ensemble of the template velocity curves 
for the eleven RRLs (shaded area). The solid and dashed purple lines show the cluster average velocity
with the errors on the mean.
It is clear that a correct phasing of the observation is fundamental to obtain more precise results.

\begin{deluxetable*}{ccccccc}
\tablecaption{Instantaneous radial velocity, systemic velocity from template and estimated stellar parameters. \label{tab:obs_param}}
\tablewidth{0pt} 
\tablehead{
\colhead{ID} &
\colhead{V$_{\text{rad}}$} &
\colhead{V$_{\gamma}$} &
\colhead{T$_{\text{eff}}$} &
\colhead{log $g$} &
\colhead{$\xi_{\text{turb}}$} &
\colhead{[M/H]} \\
\colhead{ } &
\colhead{(km~s$^{-1}$)} &
\colhead{(km~s$^{-1}$)} &
\colhead{(K)} &
\colhead{(cgs)} &
\colhead{(km~s$^{-1}$)} &
\colhead{(dex)}
}
\startdata
V3          &  497.2  $\pm$ 0.4  &   482 $\pm$ 8  &  6400  $\pm$ 100   & 2.5  $\pm$ 0.2   & 3.6  $\pm$ 0.2  &  -1.4 $\pm$ 0.1  \\
V6          &  520.5  $\pm$ 0.8  &   497 $\pm$ 8 &  6800  $\pm$ 300   & 3.3  $\pm$ 0.4   & 2.8  $\pm$ 0.4  &  -1.2 $\pm$ 0.1  \\
V14         &  463.5  $\pm$ 1.2  &   493 $\pm$ 8 &  7500  $\pm$ 300   & 3.1  $\pm$ 0.4   & 3.0  $\pm$ 0.4  &  -1.5 $\pm$ 0.1  \\
V26         &  504.1  $\pm$ 0.3  &   497 $\pm$ 8 &  6000  $\pm$ 100   & 2.1  $\pm$ 0.1   & 3.5  $\pm$ 0.1  &  -1.8 $\pm$ 0.1  \\ 
V37         &  468.4  $\pm$ 0.7  &   477 $\pm$ 8 &  7200  $\pm$ 300   & 2.5  $\pm$ 0.2   & 2.6  $\pm$ 0.4  &  -1.5 $\pm$ 0.1  \\
V38         &  504.6  $\pm$ 0.4  &   494 $\pm$ 8 &  6400  $\pm$ 200   & 2.8  $\pm$ 0.2   & 2.8  $\pm$ 0.3  &  -1.5 $\pm$ 0.1  \\
V41         &  490.1  $\pm$ 0.3  &   505 $\pm$ 8 &  6350  $\pm$ 100   & 2.3  $\pm$ 0.1   & 3.0  $\pm$ 0.1  &  -1.5 $\pm$ 0.1  \\
V47         &  484.2  $\pm$ 0.4  &   503 $\pm$ 8 &  6700  $\pm$ 200   & 2.2  $\pm$ 0.3   & 2.6  $\pm$ 0.4  &  -1.6 $\pm$ 0.1  \\
V57         &  508.5  $\pm$ 0.4  &   497 $\pm$ 8 &  6500  $\pm$ 150   & 3.0  $\pm$ 0.2   & 3.5  $\pm$ 0.3  &  -1.4 $\pm$ 0.1  \\
V73         &  467.7  $\pm$ 0.8  &   495 $\pm$ 8 &  7300  $\pm$ 150   & 2.6  $\pm$ 0.2   & 3.4  $\pm$ 0.3  &  -1.6 $\pm$ 0.1  \\  
V83         &  464.9  $\pm$ 1.0  &   492 $\pm$ 8 &  7300  $\pm$ 150   & 2.7  $\pm$ 0.2   & 2.8  $\pm$ 0.2  &  -1.5 $\pm$ 0.1  \\
94180       &  498.1  $\pm$ 0.3  &   498 $\pm$ 8 &  5600  $\pm$ 150   & 2.9  $\pm$ 0.3   & 2.2  $\pm$ 0.3  &  -1.3 $\pm$ 0.1  \\
\enddata
\end{deluxetable*}

\section{Abundance analysis} \label{sec:abund}

Metallicies and relative abundances were derived from 
equivalent widths (EW) of selected atomic transitions in our spectra.
We prepared the atomic line list selecting only the isolated, unblended lines. 
We then measured their EW by mean of a multi-gaussian fitting performed with pyEW by M. Adamow\footnote{
\url {https://github.com/madamow/pyEW}} 
and visually inspected them. 
We discarded all the highly asymmetric lines
and the ones too weak (EW$\le$10 m\AA) or too strong 
(EW$\ge$180 m\AA).
Weak lines
could be confused within the noise and their measurement errors
are too large, whereas strong lines are on the damping portion of the 
curve of growth 
and we expected larger errors associated with the retrieved abundances.
We ended up with the line list in Table~\ref{tab:righe2}.

\begin{deluxetable*}{ccccc|ccccc|ccccc}
\tablecaption{Line list and atomic parameteres. \label{tab:righe2}}
\tablewidth{0pt}
\tablehead{
\colhead{$\lambda$} &	
\colhead{Species} &
\colhead{EP} &
\colhead{log(gf)} &
\colhead{ } &
\colhead{$\lambda$} &	
\colhead{Species} &
\colhead{EP} &
\colhead{log(gf)} &
\colhead{ } &
\colhead{$\lambda$} &	
\colhead{Species} &
\colhead{EP} &
\colhead{log(gf)}  \\
\colhead{(\AA)} &	
\colhead{ } &
\colhead{(eV)} &
\colhead{(dex)} &
\colhead{ } &
\colhead{(\AA)} &	
\colhead{ } &
\colhead{(eV)} &
\colhead{(dex)} &
\colhead{ } &
\colhead{(\AA)} &	
\colhead{ } &
\colhead{(eV)} &
\colhead{(dex)}
}
\startdata
4702.991	 & Mg {\scriptsize  I} 	 & 4.346	 & -0.44 	 & &       4690.138	 & Fe {\scriptsize  I} 	 & 3.684	 & -1.68         & &      5195.472	 & Fe {\scriptsize  I} 	 & 4.217	 & 0.02 	 \\
5265.556	 & Ca {\scriptsize  I} 	 & 2.521	 & -0.26 	 & &       4728.546	 & Fe {\scriptsize  I} 	 & 3.651	 & -1.28         & &      5198.711	 & Fe {\scriptsize  I} 	 & 2.221	 & -2.09         \\
4670.407	 & Sc {\scriptsize  II}	 & 1.357	 & -0.58 	 & &       4733.591	 & Fe {\scriptsize  I} 	 & 1.484	 & -2.99 	 & &      5215.181	 & Fe {\scriptsize  I} 	 & 3.263	 & -0.86         \\
5031.021	 & Sc {\scriptsize  II}	 & 1.357	 & -0.40 	 & &       4736.773	 & Fe {\scriptsize  I} 	 & 3.209	 & -0.67 	 & &      5217.389	 & Fe {\scriptsize  I} 	 & 3.209	 & -1.07         \\
5239.813	 & Sc {\scriptsize  II}	 & 1.455	 & -0.77 	 & &       4741.529	 & Fe {\scriptsize  I} 	 & 2.829	 & -2.00 	 & &      5232.940	 & Fe {\scriptsize  I} 	 & 2.938	 & -0.19         \\
5039.957	 & Ti {\scriptsize  I} 	 & 0.021	 & -1.08 	 & &       4745.800	 & Fe {\scriptsize  I} 	 & 3.651	 & -1.25 	 & &      5242.491	 & Fe {\scriptsize  I} 	 & 3.632	 & -0.84         \\
5064.653	 & Ti {\scriptsize  I} 	 & 0.048	 & -0.94 	 & &       4786.807	 & Fe {\scriptsize  I} 	 & 3.015	 & -1.59 	 & &      5243.776	 & Fe {\scriptsize  I} 	 & 4.253	 & -1.15         \\
5210.384	 & Ti {\scriptsize  I} 	 & 0.048	 & -0.82 	 & &       4788.757	 & Fe {\scriptsize  I} 	 & 3.234	 & -1.81 	 & &      5269.537	 & Fe {\scriptsize  I} 	 & 0.858	 & -1.33         \\
4708.663	 & Ti {\scriptsize  II}	 & 1.236	 & -2.35 	 & &       4938.814	 & Fe {\scriptsize  I} 	 & 2.873	 & -1.08 	 & &      5292.597	 & Fe {\scriptsize  I} 	 & 4.987	 & -0.03         \\
4874.009	 & Ti {\scriptsize  II}	 & 3.092	 & -0.86 	 & &       4939.687	 & Fe {\scriptsize  I} 	 & 0.858	 & -3.25 	 & &      5302.303	 & Fe {\scriptsize  I} 	 & 3.281	 & -0.73         \\
4911.194	 & Ti {\scriptsize  II}	 & 3.121	 & -0.64 	 & &       4967.897	 & Fe {\scriptsize  I} 	 & 4.188	 & -0.53         & &      5324.179	 & Fe {\scriptsize  I} 	 & 3.211	 & -0.11         \\
5072.286	 & Ti {\scriptsize  II}	 & 3.121	 & -1.02 	 & &       4973.102	 & Fe {\scriptsize  I} 	 & 3.960	 & -0.69         & &      4620.513	 & Fe {\scriptsize  II}	 & 2.828	 & -3.19 	 \\
5129.156	 & Ti {\scriptsize  II}	 & 1.890	 & -1.34 	 & &       4983.250	 & Fe {\scriptsize  I} 	 & 4.151	 & -0.11         & &      4731.439	 & Fe {\scriptsize  II}	 & 2.891	 & -3.10 	 \\
5211.530	 & Ti {\scriptsize  II}	 & 2.588	 & -1.41 	 & &       5001.864	 & Fe {\scriptsize  I} 	 & 3.882	 & -0.01         & &      4993.355	 & Fe {\scriptsize  II}	 & 2.807	 & -3.70 	 \\
4600.749	 & Cr {\scriptsize  I} 	 & 1.003	 & -1.25 	 & &        5005.712	 & Fe {\scriptsize  I} 	 & 3.884	 & -0.12         & &      5197.568	 & Fe {\scriptsize  II}	 & 3.230	 & -2.05 	 \\
4616.124	 & Cr {\scriptsize  I} 	 & 0.982	 & -1.19 	 & &       5014.943	 & Fe {\scriptsize  I} 	 & 3.940	 & -0.18         & &      5234.624	 & Fe {\scriptsize  II}	 & 3.221	 & -2.21 	 \\
4626.173	 & Cr {\scriptsize  I} 	 & 0.968	 & -1.33 	 & &       5022.236	 & Fe {\scriptsize  I} 	 & 3.984	 & -0.33         & &      5264.801	 & Fe {\scriptsize  II}	 & 3.230	 & -3.23 	 \\
4646.162	 & Cr {\scriptsize  I} 	 & 1.029	 & -0.74 	 & &       5044.211	 & Fe {\scriptsize  I} 	 & 2.849	 & -2.15 	 & &      5284.092	 & Fe {\scriptsize  II}	 & 2.891	 & -3.20 	 \\
4651.291	 & Cr {\scriptsize  I} 	 & 0.982	 & -1.46 	 & &       5049.819	 & Fe {\scriptsize  I} 	 & 2.277	 & -1.35 	 & &      4648.652	 & Ni {\scriptsize  I} 	 & 3.417	 & -0.09 	 \\
4652.157	 & Cr {\scriptsize  I} 	 & 1.003	 & -1.04 	 & &       5074.748	 & Fe {\scriptsize  I} 	 & 4.217	 & -0.20         & &      4866.271	 & Ni {\scriptsize  I} 	 & 3.536	 & -0.22 	 \\
5296.691	 & Cr {\scriptsize  I} 	 & 0.982	 & -1.36 	 & &       5083.339	 & Fe {\scriptsize  I} 	 & 0.957	 & -2.84 	 & &      5035.362	 & Ni {\scriptsize  I} 	 & 3.633	 & 0.29 	 \\
4616.629	 & Cr {\scriptsize  II}	 & 4.069	 & -1.29 	 & &       5090.773	 & Fe {\scriptsize  I} 	 & 4.253	 & -0.40         & &      5081.107	 & Ni {\scriptsize  I} 	 & 3.844	 & 0.30 	 \\
4634.073	 & Cr {\scriptsize  II}	 & 4.069	 & -0.98 	 & &       5123.720	 & Fe {\scriptsize  I} 	 & 1.010	 & -3.06 	 & &      5084.089	 & Ni {\scriptsize  I} 	 & 3.676	 & 0.03 	 \\
5237.329	 & Cr {\scriptsize  II}	 & 4.070	 & -1.16 	 & &       5127.360	 & Fe {\scriptsize  I} 	 & 0.914	 & -3.25 	 & &      5099.927	 & Ni {\scriptsize  I} 	 & 3.676	 & -0.10 	 \\
5308.408	 & Cr {\scriptsize  II}	 & 4.068	 & -1.81 	 & &       5131.468	 & Fe {\scriptsize  I} 	 & 2.221	 & -2.52 	 & &      5115.389	 & Ni {\scriptsize  I} 	 & 3.831	 & -0.11 	 \\
5313.563	 & Cr {\scriptsize  II}	 & 4.070	 & -1.65 	 & &       5133.689	 & Fe {\scriptsize  I} 	 & 4.175	 & 0.36 	 & &      5176.559	 & Ni {\scriptsize  I} 	 & 3.895	 & -0.44 	 \\
4598.117	 & Fe {\scriptsize  I} 	 & 3.281	 & -1.57 	 & &       5141.739	 & Fe {\scriptsize  I} 	 & 2.422	 & -2.15         & &      4722.153	 & Zn {\scriptsize  I} 	 & 4.030	 & -0.33 	 \\
4602.000	 & Fe {\scriptsize  I} 	 & 1.607	 & -3.13 	 & &       5150.840	 & Fe {\scriptsize  I} 	 & 0.989	 & -3.04         & &      4810.528	 & Zn {\scriptsize  I} 	 & 4.078	 & -0.14 	 \\
4602.941	 & Fe {\scriptsize  I} 	 & 1.484	 & -2.21 	 & &       5151.911	 & Fe {\scriptsize  I} 	 & 1.010	 & -3.32         & &      4883.684	 &  Y {\scriptsize  II}	 & 1.083	 & 0.07 	 \\
4619.288	 & Fe {\scriptsize  I} 	 & 3.600	 & -1.06 	 & &       5159.058	 & Fe {\scriptsize  I} 	 & 4.280	 & -0.82 	 & &      4900.110	 &  Y {\scriptsize  II}	 & 1.032	 & -0.09 	 \\
4625.045	 & Fe {\scriptsize  I} 	 & 3.239	 & -1.27 	 & &       5162.273	 & Fe {\scriptsize  I} 	 & 4.175	 & 0.02  	 & &      5087.420	 &  Y {\scriptsize  II}	 & 1.083	 & -0.17 	 \\
4678.846	 & Fe {\scriptsize  I} 	 & 3.600	 & -0.68 	 & &       5194.942	 & Fe {\scriptsize  I} 	 & 1.556	 & -2.02  	 & &      5200.413	 &  Y {\scriptsize  II}	 & 0.992	 & -0.57 	 \\
\enddata
\tablenotetext{}{Sources:
Mg {\scriptsize  I}, NIST database \citep{kramida18}; 
Ca {\scriptsize  I}, NIST; 
Sc {\scriptsize  II}, NIST; 
Ti {\scriptsize  I}, \cite{lawler13}; 
Ti {\scriptsize  II}, \cite{wood13}; 
Cr {\scriptsize  I}, \cite{sobeck07}; 
Cr {\scriptsize  II}, \cite{lawler17};
Fe {\scriptsize  I}, \cite{obrian91,denhartog14,ruffoni14,belmonte17}; 
Fe {\scriptsize  II}, NIST;
Ni {\scriptsize  I}, \cite{wood14};
Zn {\scriptsize  I},  VALD database \citep{ryabchikova15};
Y {\scriptsize  II}, NIST.} %\cite{biemont11}.}
%\tablenotetext{a}{\url{https://www.nist.gov/pml/atomic-spectra-database}}
%\tablenotetext{b}{\url{http://vald.astro.univie.ac.at/~vald3/php/vald.php}}
\end{deluxetable*}

\begin{figure}
\plotone{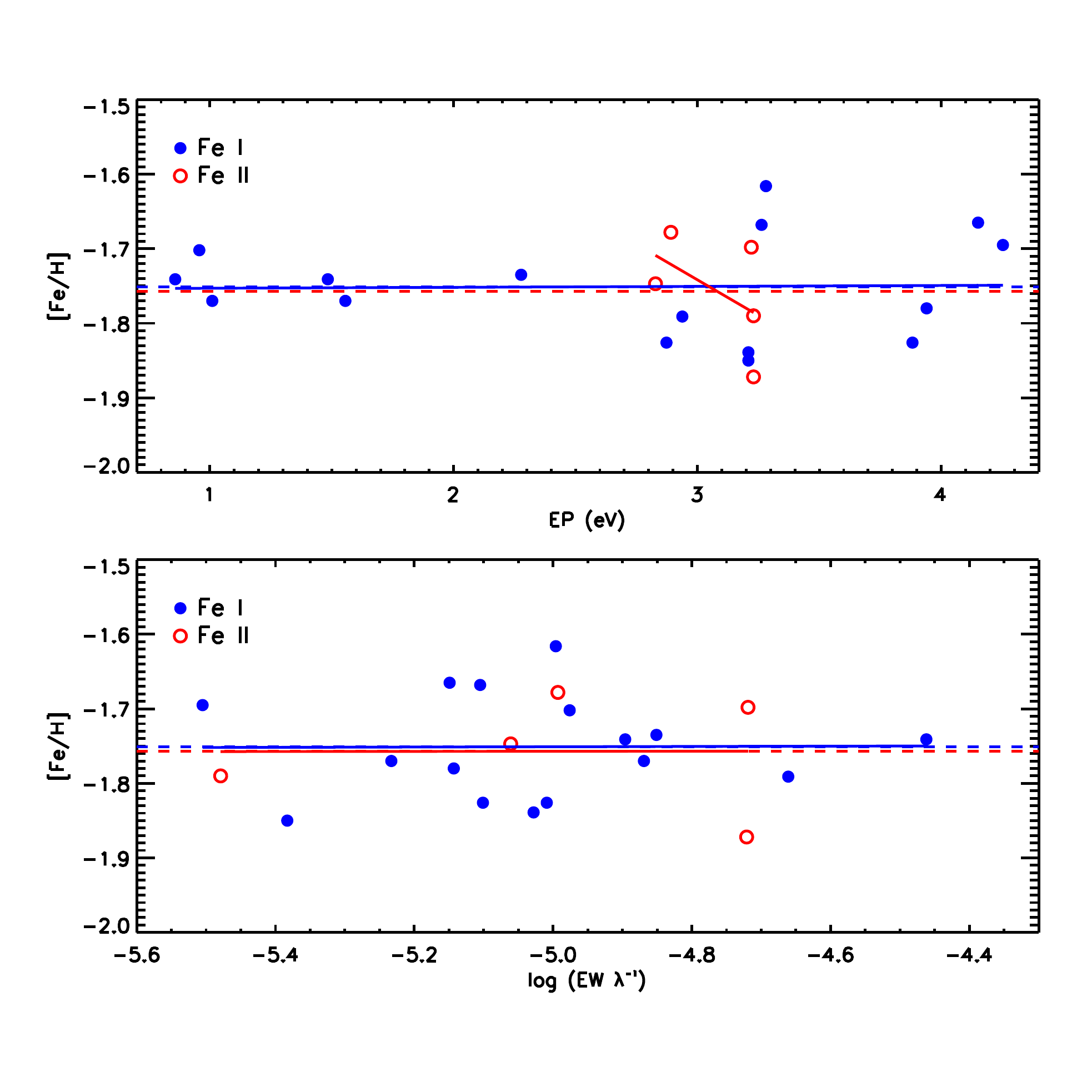}
\caption{Individual iron line abundances for V26 in NGC~3201 as a function of excitation potential (top panel) and reduced equivalent width (bottom panel). Average abundances are shown with dashed lines. Linear fits of data points are shown with solid lines. Linear trends are minimized and neutral/ionized species are balanced to equilibrium. \label{fig:moog}}
\end{figure}

We used the LTE line analysis code MOOG \citep{sneden73}, implemented in the 
Python wrapper pyMOOGi\footnote{
\url {https://github.com/madamow/pymoogi}}
\citep{adamow17},
and a grid of $\alpha$-enhanced models from \cite{castelli03}\footnote{
\url{http://kurucz.harvard.edu/grids.html}}
to estimate the atmospheric parameters (\teff, \logg, \vmicro) 
and the abundances for the sample spectra.
The entire analysis mainly concentrated on Fe~I and Fe~II lines to 
estimate the proper 
parameters; the final set for each star is given in 
Table~\ref{tab:obs_param}.
We searched for the spectroscopically defined parameters set following a 
standard approach. 
An example is shown in Figure~\ref{fig:moog}. 

i) The effective temperature is estimated in such a way that abundances 
from individual lines do not show dependence on the excitation potential
(EP, top panel of the figure). 

ii) The microturbulence is estimated limiting the dependence of the 
individual line abundances on the reduced equivalent width ($\log$(EW~$\lambda^{-1}$), 
bottom panel of the figure).

iii) The surface gravity is estimated assuming the balance between the 
ionization states, minimizing the differences between neutral and ionized 
species.

\begin{figure*}
\plotone{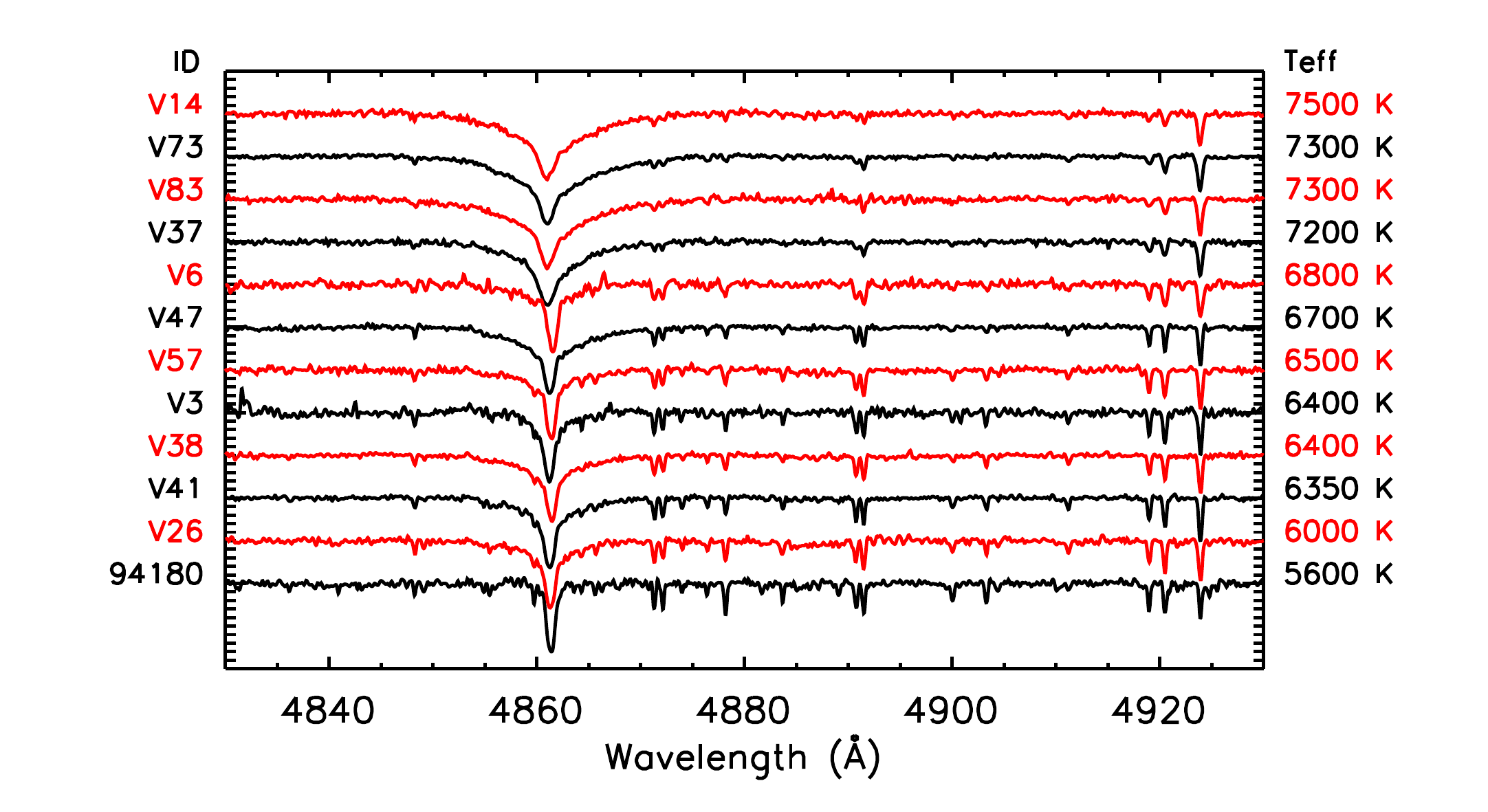}
\caption{Portion of the observed M2FS spectra for the analysed stars (listed on the left), sorted by estimated temperatures (listed on the right). The bottom spectrum is the cooler RHB star. \label{fig:teff}}
\end{figure*}

\begin{figure}
\plotone{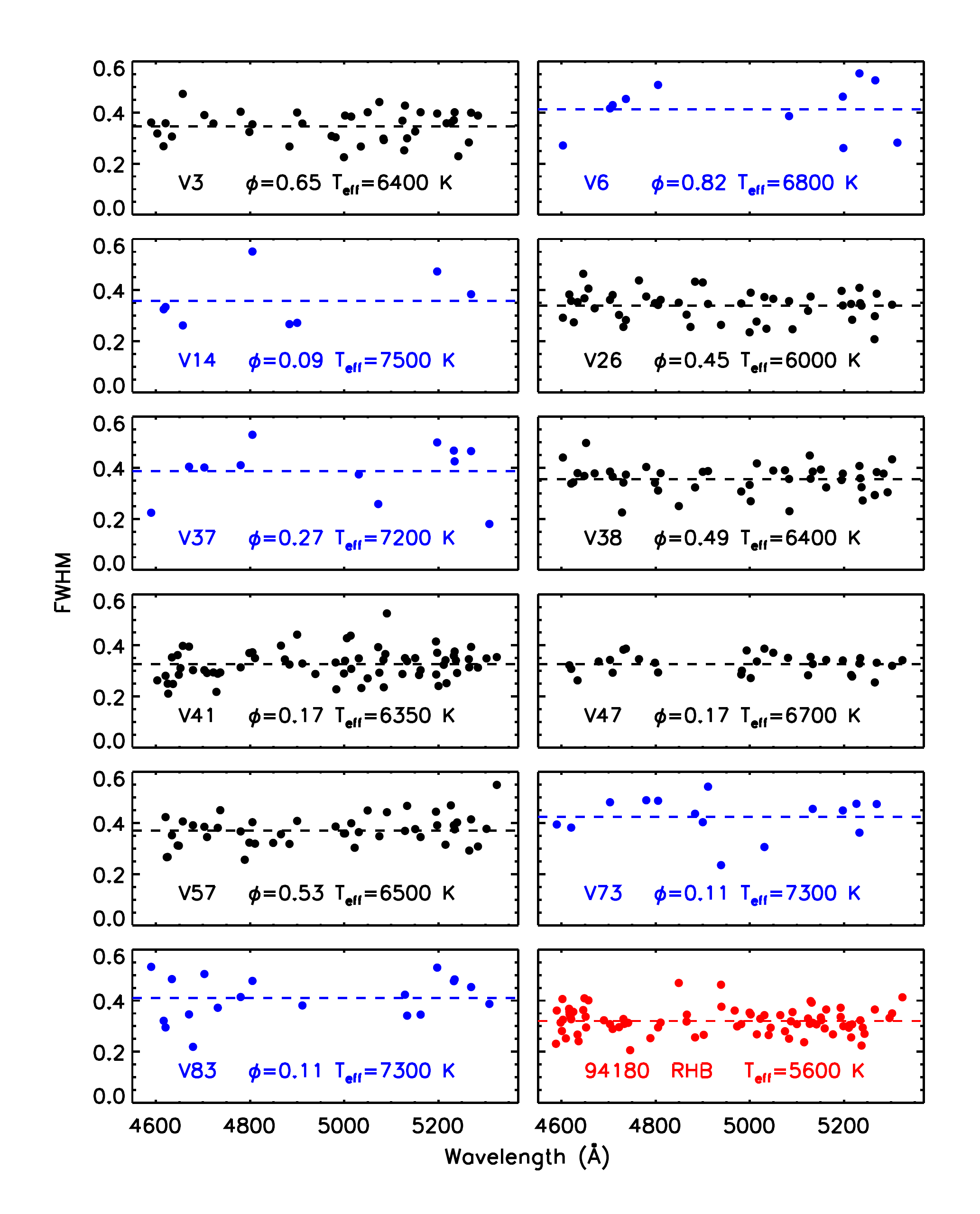}
\caption{Individual FWHM of all the measured lines in the sample stars. Blue color highlights the warmest stars. Red color highlights the coolest, RHB star. \label{fig:fwhm}}
\end{figure}

The EW analysis is bolstered by two other exercises with
our spectra.
In Figure~\ref{fig:teff} we show a montage of the spectra in the wavelength
region 4830--4930~\AA, ordering the stars by their derived temperatures.
The dependence on \teff\ appears: the warmer stars have weaker atomic
line absorptions and stronger 4861~\AA\ H$\beta$ lines.
In Figure~\ref{fig:fwhm} we show the FWHM values for all the measured atomic lines of
each star, highlighting with blue color the five RR~Lyraes with highest 
derived \teff\ values.
The mean FWHM values range from 0.33 to 0.43~\AA, larger than that which 
would be generated by a combination of spectrograph, thermal, and microturbulent broadening.  
We attribute this extra width to a combination of macroturbulent and 
possible rotational broadening.  Upper limits to axial rotation of RRab field 
stars is estimated to be $\sim$5~\kmsec\ \citep{preston18},
These stars, in comparison with the six cooler RR~Lyraes, have three common 
characteristics: far fewer measurable lines 
($\langle n_{\rm warm}\rangle$~$\simeq$~13 and
$\langle n_{\rm cool}\rangle$~$\simeq$~46);
larger FWHM values 
($\langle FWHM_{\rm warm}\rangle$~$\simeq$~0.40 and
$\langle FWHM_{\rm cool}\rangle$~$\simeq$~0.35);
and phases closer to maximum light
($\langle \phi_{\rm warm}\rangle$~$\simeq$0.08,
$\langle \phi_{\rm cool}\rangle$~$\simeq$0.41);
% V6, V14, V37, V73, V83 hot
% V3, V26, V38, V41, V47, V57 cool
% 94180, called V81 previously, RHB

Some variation on the EW procedure was applied
for V14.
Only one Fe I line is observed in its spectrum, so that
it is impossible to estimate the effective temperature 
as described above.
However, a visual inspection of its spectrum
clearly shows that the few observable lines are weaker than
the counterparts in the other stars, 
suggesting that this star is the hottest one in the sample.
Based on this visual evidence we estimated \teff~$\sim$~7500~K for V14.

We also call attention to star 94180\footnote{The star is 
identified as 94180 in our catalogue of NGC~3201 stars, 
provided by P. B. Stetson (private communication). SIMBAD
online catalogue identifies it as NGC~3201~CWFD~3-327. See Table~\ref{tab:position} for more details.}, the NGC~3201 RHB 
star observed along with the RR~Lyrae sample.
Our Fe EW analysis clearly indicated a temperature (\teff~=~5600~K) much 
cooler than the red edge of the RR~Lyrae instability strip.
Inspection of Figure~\ref{fig:teff} confirms the EW analysis.
This star has the deepest atomic lines, and its H$\beta$ line has 
weak-to-absent damping wings.
The FWHM data for 94180 (Figure~\ref{fig:fwhm}) are in accord:  this
star has the largest number of measured lines (81) and the smallest
measured line widths ($\langle FWHM\rangle$~$\simeq$~0.32) of all of
our stars.

\begin{figure}
\plotone{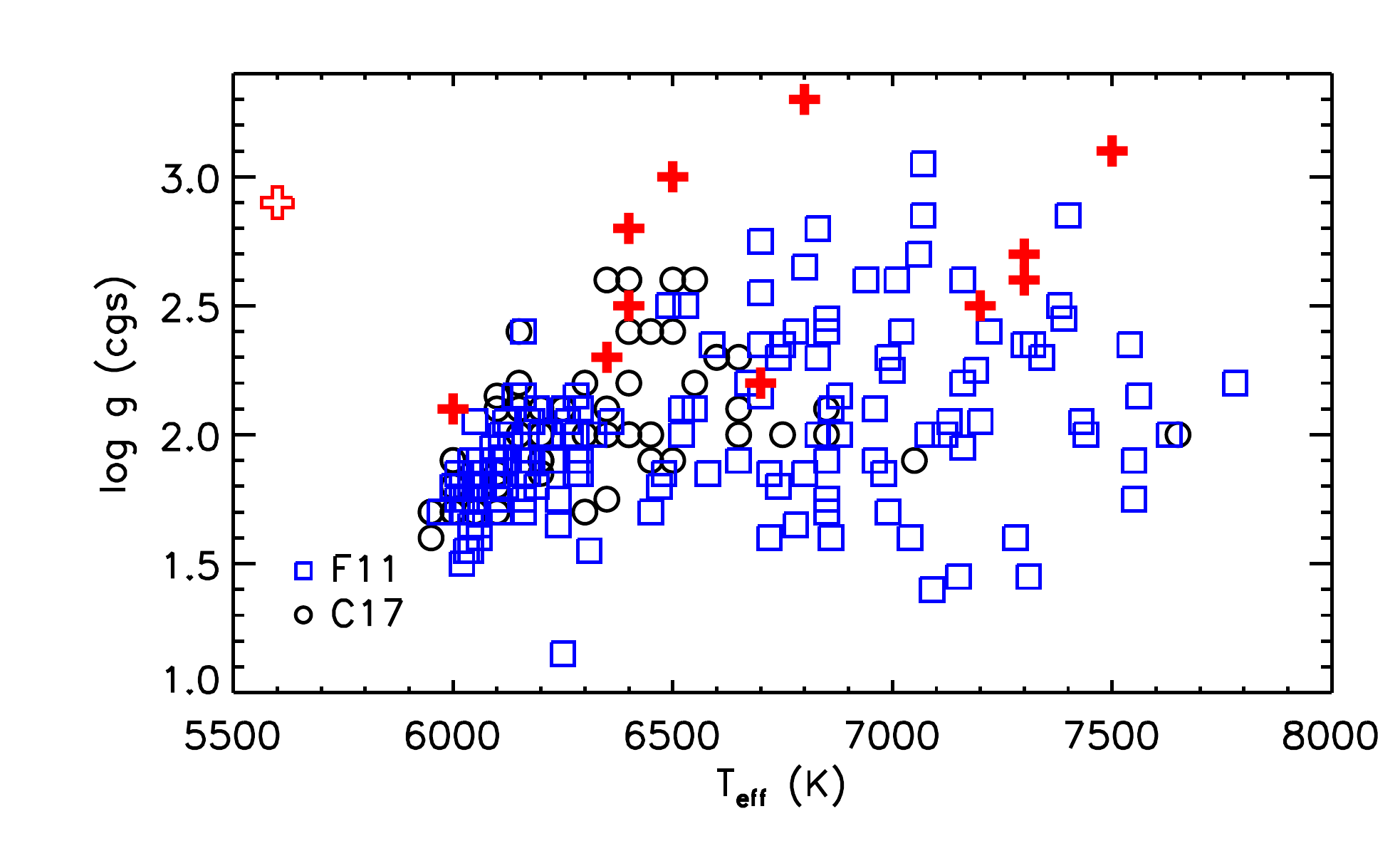}
\caption{\teff\ vs \logg\ for the stars in our sample (RRLs: filled red crosses; RHB: empty red cross) and a sample of field RRLs available in the literature (F11=\citealt{for11}; C17=\citealt{chadid17}). \label{fig:tefflogg}}
\end{figure}

Finally, in Figure~\ref{fig:tefflogg} we compare estimated effective 
temperatures and surface gravities of our sample with those from a couple of
literature spectroscopic studies of field RR Lyrae stars. 
Our RRLs appear to have systematically higher gravities
than the field RRLs, suggesting the latter to be slightly more
evolved than the cluster stars. Moreover, the difference
between the RRLs and the RHB star is evident.

We estimated the internal errors in iron abundances associated 
with changes in the model parameters
by varying effective temperature, surface gravity and microturbulence 
in the line analysis of V41,
selected as a representative of the entire cluster.
Variations in steps of $\Delta$T$_{\text{eff}}$=100~K, $\Delta$log $g$=0.3, 
$\Delta\xi_{turb}$=0.5~km~s$^{-1}$ were applied, considering them as 
independent parameters.
Results are shown in Table~\ref{tab:err_param}.
Effective temperature and surface gravity are the main sources of 
error for Fe~I and Fe~II abundances, respectively,
whereas the impact of microturbulence is on average a factor of two smaller. 
The total error associated with the parameters (last column in 
Table~\ref{tab:err_param}) is then computed by adding in quadrature the 
individual errors on the intrinsic parameters.
The final estimated Fe~I and Fe~II abundance uncertainties
due to model atmosphere uncertainties are about 0.01~dex for each species.

\begin{deluxetable}{lcccc} 
\tablecaption{Errors on iron abundances associated with errors on the paramater estimates. \label{tab:err_param}} 
\tablewidth{0pt} 
\tablehead{ 
\colhead{Species} & 
\colhead{$\Delta$\teff} & 
\colhead{$\Delta$\logg} & 
\colhead{$\Delta$\vmicro} & 
\colhead{$\sigma_{\text{tot}}$}\\ 
\colhead{ } & 
\colhead{(100 K)} & 
\colhead{(0.3 dex)} & 
\colhead{(0.5 km s$^{-1}$)} & 
\colhead{ } 
} 
\startdata 
$\Delta$[FeI/H]  & 0.07 & 0.01 & 0.04 & 0.08 \\ 
$\Delta$[FeII/H] & 0.02 & 0.10 & 0.06 & 0.12 \\ 
\enddata 
\end{deluxetable}

A comprehensive abundance analysis cannot leave possible NLTE corrections
out of consideration.  
Several works are available in recent literature dealing 
with NLTE in RRL stars \citep{wallerstein10,hansen11,andrievsky18}, 
but the analysis is far to be complete. 
Indeed, not all the elements we are dealing with have already been studied.
NLTE corrections can affect the abundances up to $\sim$0.5~dex \citep{hansen11}
but they are strongly dependent on temperature and the details of the
calculations.
Additionally, the M2FS wavelength coverage limits the number of useful
transitions available to make a serious study of NLTE effects.
For these reasons, we decided to take into account only LTE effects, 
obtaining reasonable metallicities and abundance ratios compared with RGB stars (see next sections).

\subsection{Iron Metallicity} \label{sec:iron}

We estimated the average iron abundance of NGC~3201 to be
$\langle$[Fe/H]$\rangle$ = $-1.47 \pm 0.04$, with a dispersion 
$\sigma$ = 0.14 dex.
Table~\ref{tab:fe} lists the individual iron abundances for the current 
sample with the intrinsic errors quantifying the line to line 
variability. 
Data listed in this table indicate that our mean [Fe/H] 
estimate is compatible with a homogeneous, mono-metallic cluster.
Note that we are dealing with 
variable stars and once 
uncertainties in the intrinsic parameters are taken into 
account, the mono-metallicity of the cluster 
is further supported.

\begin{deluxetable*}{crlcrlcrl}
\tablecaption{Iron abundances and number of lines for each ionization state. \label{tab:fe}}
\tablewidth{0pt}
\tablehead{
\colhead{ID} &
\twocolhead{[FeI/H]} &
\colhead{n} &
\twocolhead{[FeII/H]} &
\colhead{n} &
\twocolhead{[Fe/H]}
}
\startdata
V3       &    -1.37  & $\pm$    0.05  &     16  &    -1.36  & $\pm$    0.03  &      5  &  -1.37  &  $\pm$  0.03  \\ 
V6       &    -1.18  & $\pm$    0.04  &      3  &    -1.22   &                &      1  &  -1.19  &  $\pm$  0.03  \\ 
V14      &    -1.57  &                &      1  &    -1.50  & $\pm$    0.16  &      2  &  -1.53  &  $\pm$  0.09  \\ 
V26      &    -1.75  & $\pm$    0.02  &     16  &    -1.76  & $\pm$    0.03  &      5  &  -1.75  &  $\pm$  0.02  \\ 
V37      &    -1.49  & $\pm$    0.06  &      2  &    -1.52  & $\pm$    0.04  &      2  &  -1.51  &  $\pm$  0.03  \\ 
V38      &    -1.48  & $\pm$    0.04  &     18  &    -1.49  & $\pm$    0.04  &      5  &  -1.48  &  $\pm$  0.03  \\ 
V41      &    -1.50  & $\pm$    0.03  &     28  &    -1.48  & $\pm$    0.04  &      6  &  -1.50  &  $\pm$  0.03  \\ 
V47      &    -1.60  & $\pm$    0.03  &     15  &    -1.59  & $\pm$    0.08  &      5  &  -1.60  &  $\pm$  0.03  \\ 
V57      &    -1.35  & $\pm$    0.04  &     20  &    -1.36  & $\pm$    0.05  &      5  &  -1.35  &  $\pm$  0.03  \\ 
V73      &    -1.55  & $\pm$    0.03  &      4  &    -1.55  & $\pm$    0.10  &      2  &  -1.55  &  $\pm$  0.03  \\ 
V83      &    -1.48  & $\pm$    0.08  &      5  &    -1.41  & $\pm$    0.10  &      4  &  -1.45  &  $\pm$  0.06  \\ 
94180    &    -1.33  & $\pm$    0.03  &     37  &    -1.32  & $\pm$    0.07  &      3  &  -1.33  &  $\pm$  0.03  \\ 
\hline
NGC 3201 &    -1.47  & $\pm$    0.04  &         &    -1.46  & $\pm$    0.04  &         &  -1.47  &  $\pm$  0.04  \\ 
\enddata
\end{deluxetable*}

\begin{figure}
\plotone{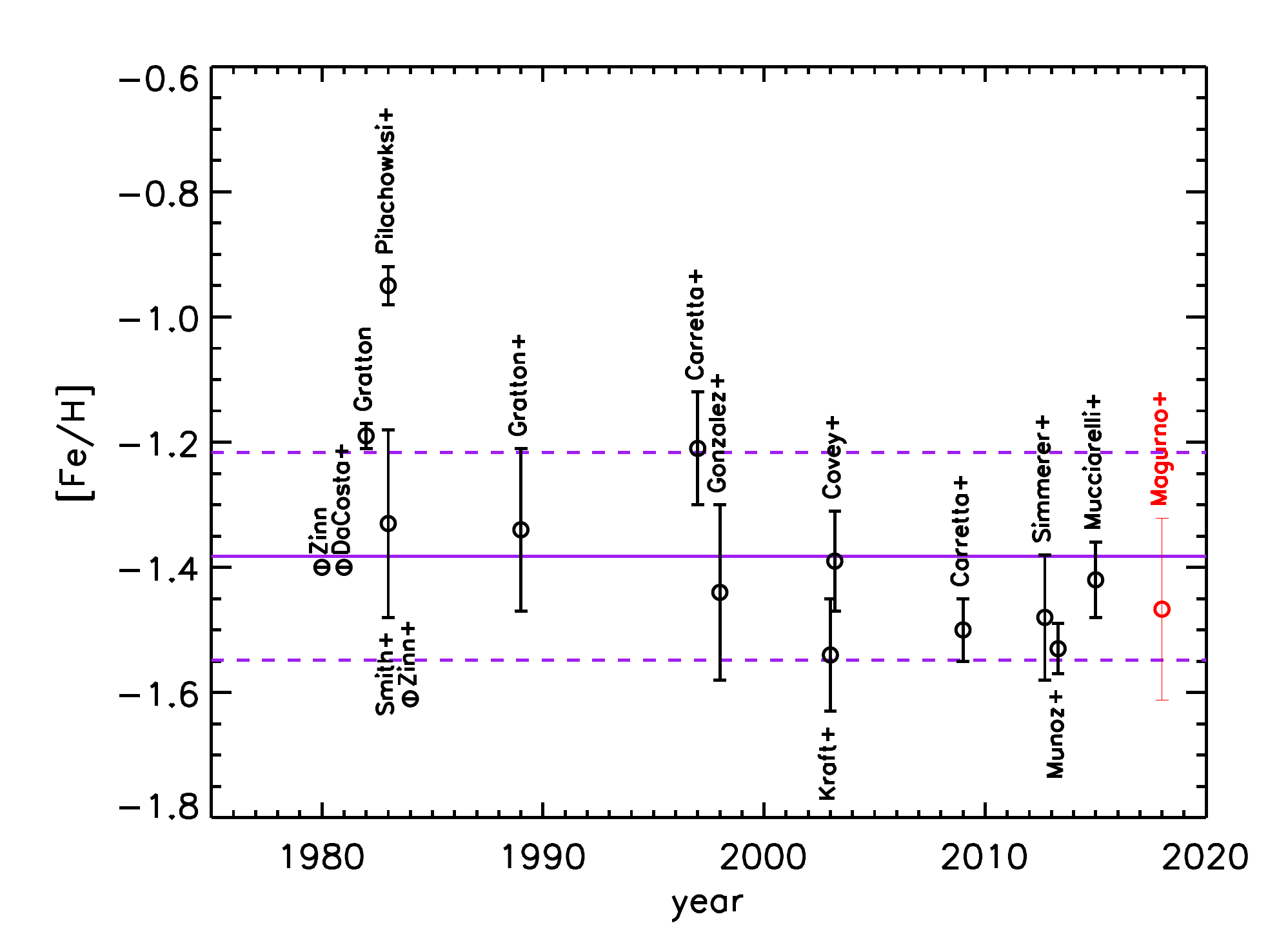}
\caption{Literature determination of [Fe/H] for NGC~3201 \citep[][see also Table~\ref{tab:literature}]{zinn80,dacosta81,gratton82,pilachowski83,smith83,zinn84,gratton89,carretta97,gonzalez98,covey03,kraft03,carretta09,munoz13,simmerer13,mucciarelli15}. Error bars show the intrinsic dispersions. The sample mean and one sigma dispersion are shown with purple lines.  \label{fig:fehistory}}
\end{figure}

NGC~3201 has been at the cross-road of several spectroscopic investigations 
in the recent literature (see Table~\ref{tab:literature} for a detailed list).
Figure~\ref{fig:fehistory} 
shows several published iron abundance determinations
for NGC~3201, and error bars display the standard deviations ($\sigma$) 
of the different samples.
The solid and dashed purple lines show the average and 1$\sigma$ 
dispersion of the entire sample, compared with our result.
The data plotted in the figure show that 
the bulk of the iron abundances for NGC~3201 do agree within 1$\sigma$.   

\begin{deluxetable}{rlccc}
\tablecaption{Literature iron abundances and number of analysed stars in NGC~3201. \label{tab:literature}}
\tablewidth{0pt}
\tablehead{
\twocolhead{[Fe/H]} &
\colhead{$\sigma$} &
\colhead{n} &
\colhead{Reference\tablenotemark{a}}
}
\startdata
-1.40 & $\pm$ 0.06        & \ldots &  \dots &          \cite{zinn80}   \\         
-1.40 & $\pm$ 0.2          & \ldots &   26   &          \cite{dacosta81}   \\      
-1.19 & $\pm$ 0.05        &  0.02  &    2   &          \cite{gratton82}   \\      
-0.95 & $\pm$ 0.2         &  0.03  &    4   &          \cite{pilachowski83} \\    
-1.33 & $\pm$ 0.05        &  0.15  &    9   &          \cite{smith83}  \\         
-1.61 & $\pm$ 0.12        & \ldots & \ldots &          \cite{zinn84} \\           
-1.34 & $\pm$ 0.05        &  0.13  &    3   &          \cite{gratton89} \\        
-1.21 & $\pm$ 0.05        &  0.09  &    3   &          \cite{carretta97} \\       
-1.44 & $\pm$ 0.03        &  0.14  &   18   &          \cite{gonzalez98} \\       
-1.39 & $\pm$ 0.06        &  0.08  &    5   &          \cite{covey03} \\          
-1.54 & $\pm$ 0.10        &  0.09  &   13   &          \cite{kraft03} \\          
-1.50 & $\pm$ 0.02        &  0.05  &  162   &          \cite{carretta09} \\       
-1.53 & $\pm$ 0.01        &  0.04  &    8   &          \cite{munoz13} \\          
-1.48 & $\pm$ 0.02        &  0.1   &   26   &          \cite{simmerer13} \\       
-1.42 & $\pm$ 0.02        &  0.06  &   21   &          \cite{mucciarelli15} \\    
\enddata
\tablenotetext{a}{Scaled to \cite{asplund09}.}
\end{deluxetable}

In particular, the most recent (ten years) estimates further suggest that 
NGC~3201 is a canonical metal-intermediate globular. 
Indeed, the average abundance for these recent works is 
$\langle$[Fe/H]$\rangle$ = $-1.48 \pm 0.02$, $\sigma = 0.05$.
Differences between the authors reflect their different approaches:

a) The sample sizes vary by two orders of magnitude,
from 2 \citep{gratton82} to 162 \citep{carretta09} stars, so that the 
final abundance might not be representative of the entire cluster in some 
cases.

b) Different techniques are used to estimate the atmospheric parameters. 
Photometric or spectroscopic estimates of effective temperature and 
surface gravity are used, based on differential colors or on line intensities.
Different techniques can lead to different results even on the same data 
sample.

c) The cluster metallicity is estimated with different methods, 
using either spectroscopic or photometric techniques 
($\Delta S$: \citealt{smith83}; $Q_{39}$: \citealt{zinn84}; 
EW: \citealt{gonzalez98}).  

Most studies generally agree on the average NGC~3201 
metallicity, but the possible existence of an intrinsic [Fe/H] spread 
within the cluster is not settled.
The definition of {\it spread} can vary among the authors, so here we 
define as {\it spread} the standard deviation $\sigma$ of the sample.
\cite{smith83}, \cite{kraft03} and \cite{munoz13} found no evidence of variation in the 
iron content of the cluster.
\cite{carretta09} analysed the largest sample of stars (162) in NGC~3201,
and also found an internal metallicity spread of only 0.05 dex.
On the other hand, \cite{gonzalez98} and \cite{simmerer13} reported
an internal [Fe/H] spread in NGC~3201 of 0.14 and 0.1 dex respectively,
with a difference as large as $\sim$0.4 dex between the 
highest and the lowest metallicities of
the cluster members.
Their abundance analyses were based on large samples of cluster members 
(18 and 26 stars). 
However, different authors using the same spectra obtained different 
conclusions about the metallicity spread in NGC~3201.
\cite{covey03} analysed a sub-sample of the \cite{gonzalez98} spectra, 
supporting a spread in iron of $\sim$0.14 dex when estimating the 
effective temperatures using spectroscopic diagnostics. 
However, the spread in iron decreased to $\sim$0.08 dex when \teff\ values
were estimated with photometric diagnostics.
A similar result was also obtained by \cite{mucciarelli15} using 
the \cite{simmerer13} spectra. 
\citeauthor{covey03} and \citeauthor{mucciarelli15} suggested that the 
spread in iron abundance shows up in spectroscopic analyses that 
do not properly take account for non-LTE effects.

The star-to-star scatter derived in our LTE analysis of 
NGC~3201 RR~Lyrae stars is small, $\sigma \sim$ 0.14, and is in accord
with prior publications that assert that this cluster is mono-metallic.

%_______________________________________________________________________________
\subsection{$\alpha$-elements: Mg, Ca, Ti} \label{sec:alpha}
%_______________________________________________________________________________

The restricted wavelength coverage of our M2FS spectra
limited the number of transitions for true $\alpha$-elements to one each for
Mg~I and Ca~I (Tables~\ref{tab:righe2} and \ref{tab:fealpha}),
not allowing a detailed analysis of these two elements.
Although Ti is not a \textquotedblleft pure\textquotedblright\ 
$\alpha$-element, because its dominant isotope is $^{48}$Ti instead of 
$^{44}$Ti, we included it in this group because its abundance at
low metallicity usually mimics those of other $\alpha$-elements.
We measured up to 17 Ti~I and Ti~II lines per spectrum in the 
best cases, so Ti abundances are the most precise among the three $\alpha$.
The mean estimated abundances are: [Mg/Fe]=0.13$\pm$0.05, 
[Ca/Fe]=0.15$\pm$0.07 and [Ti/Fe]=0.46$\pm$0.04 (Table~\ref{tab:fealpha}).

\begin{deluxetable*}{cccccrlcrl}
\tablecaption{$\alpha$-elements abundances and number of lines. \label{tab:fealpha}}
\tablewidth{0pt} 
\tablehead{
\colhead{ID} &
\colhead{[Mg/Fe]} &
\colhead{n} &
\colhead{[Ca/Fe]} &
\colhead{n} &
\twocolhead{[Ti/Fe]} &
\colhead{n} &
\twocolhead{[$\alpha$/Fe]} 
}
\startdata
V3       &      0.09   &      1  &   \nodata  & \nodata &     0.34  & $\pm$   0.07  &     10  &   \nodata  & $\pm$ \nodata  \\
V6       &     -0.09   &      1  &    -0.13   &      1  &     0.61  & $\pm$   0.18  &      2  &      0.25  & $\pm$    0.27  \\
V14      &    \nodata  & \nodata &   \nodata  & \nodata &     0.76  & $\pm$   0.12  &      2  &   \nodata  & $\pm$ \nodata  \\
V26      &      0.20   &      1  &     0.27   &      1  &     0.43  & $\pm$   0.04  &     13  &      0.40  & $\pm$    0.10  \\
V37      &      0.15   &      1  &   \nodata  & \nodata &     0.35  & $\pm$   0.08  &      4  &   \nodata  & $\pm$ \nodata  \\
V38      &      0.24   &      1  &     0.14   &      1  &     0.35  & $\pm$   0.03  &      9  &      0.32  & $\pm$    0.06  \\
V41      &      0.09   &      1  &     0.22   &      1  &     0.39  & $\pm$   0.03  &     16  &      0.37  & $\pm$    0.09  \\
V47      &      0.31   &      1  &     0.11   &      1  &     0.26  & $\pm$   0.02  &      7  &      0.25  & $\pm$    0.04  \\
V57      &      0.13   &      1  &     0.01   &      1  &     0.60  & $\pm$   0.04  &     12  &      0.54  & $\pm$    0.13  \\
V73      &      0.30   &      1  &   \nodata  & \nodata &     0.49  & $\pm$   0.04  &      5  &   \nodata  & $\pm$ \nodata  \\
V83      &      0.22   &      1  &   \nodata  & \nodata &     0.49  & $\pm$   0.11  &      5  &   \nodata  & $\pm$ \nodata  \\
94180    &     -0.19   &      1  &     0.45   &      1  &     0.46  & $\pm$   0.08  &     17  &      0.43  & $\pm$    0.20  \\
\hline
NGC 3201 & 0.13 $\pm$ 0.05   &   &  0.15 $\pm$ 0.07 &   &     0.46  & $\pm$   0.04  &         &      0.37  & $\pm$   0.04   \\       
\enddata
\end{deluxetable*}

\begin{figure}
\plotone{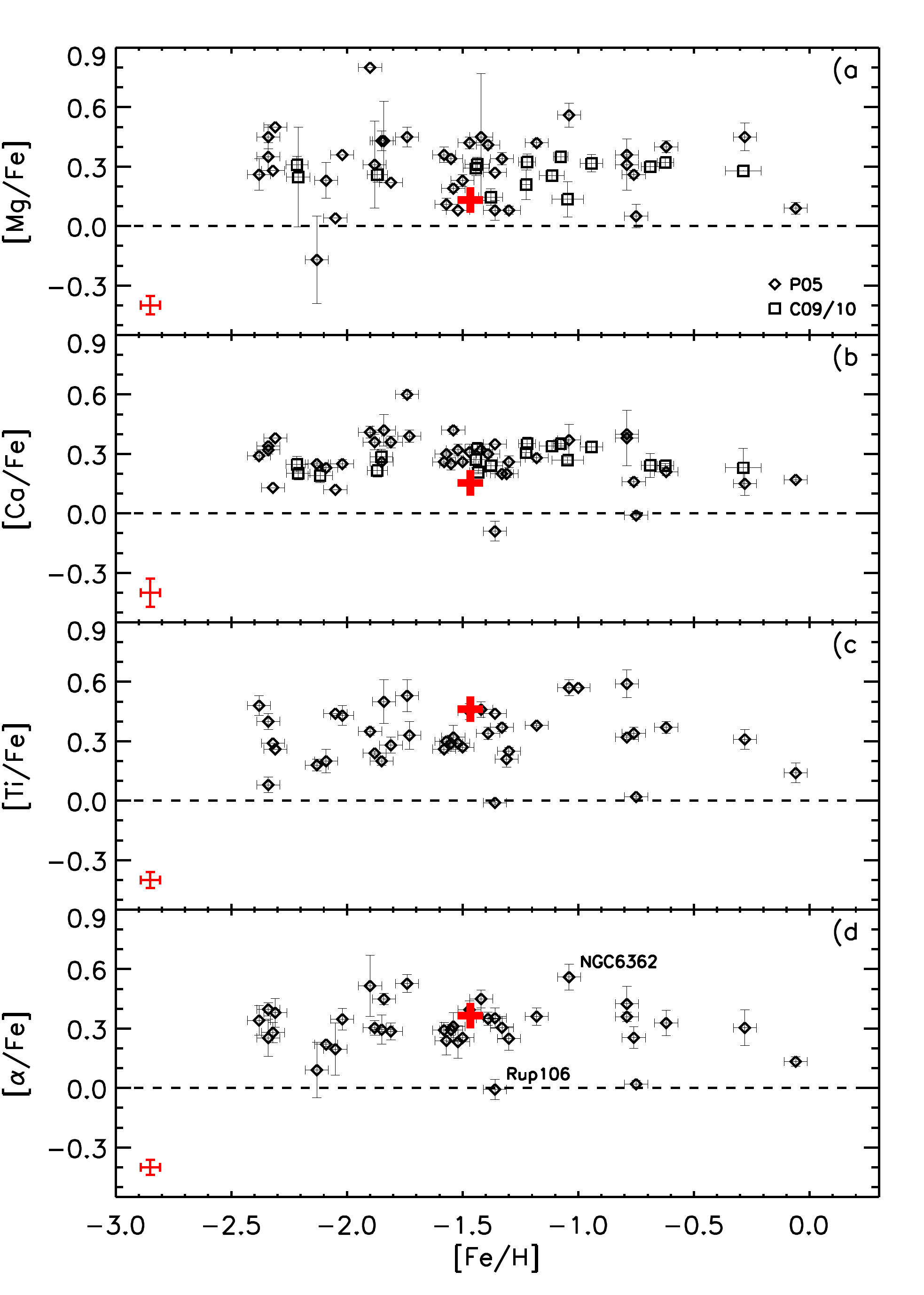}
\caption{$\alpha$-elements vs iron abundances of Galactic globular clusters (P05=\citealt{pritzl05b}; C09/10=\citealt{carretta09b,carretta10}). The red cross, with red error bar in the bottom left corner, shows our analysis of NGC~3201.  \label{fig:clustersep}}
\end{figure}

To further investigate the $\alpha$-element abundances of NGC~3201 in the context of 
Galactic globulars, Figure~\ref{fig:clustersep} shows the comparison with measurements 
available in the literature (\citealt{pritzl05b,carretta09b,carretta10}, re-scaled to the solar abundances of \citealt{asplund09}). 
We performed a linear fit over the individual $\alpha$-element abundances versus the iron content, 
in the range of metal--intermediate globulars ($-1.7 \lesssim$ [Fe/H] $\lesssim -1.0$),
to define the dispersion of the sample in the neighbourhood of NGC~3201.
In the context of Galactic globulars, Ti abundance agrees quite well within the dispersion (0.10 dex),
whereas Mg and Ca, that have respectively the highest (0.13 dex) and the lowest (0.06 dex) dispersion,
are located in the lower envelope of the observed abundance distribution, but still within 1$\sigma$. 
 
The total abundance of $\alpha$-elements (panel d in Figure \ref{fig:clustersep}) was estimated as the biweight mean
of the abundances, with respect to iron, for the three individual elements. 
Biweight is a resistant and robust estimator of location, 
more insensitive to outliers than a simple mean or median 
thanks to an iterative process (more details in \citealt{beers90}).
To perform a solid comparison with similar estimates 
available in the literature we only included the sample stars for which we were able to measure 
the three quoted $\alpha$-elements. We ended up with seven stars  and we 
found [$\alpha$/Fe]=0.37$\pm$0.04 (see Table~\ref{tab:fealpha}). On the other hand, if we 
estimate the biweight mean of the $\alpha$-elements either as Ca+Ti or as Mg+Ti we end up 
with [$\alpha$/Fe]=0.40$\pm$0.04 and  [$\alpha$/Fe]=0.39$\pm$0.03, respectively,
thus suggesting very similar enhancements. The comparison of the mean $\alpha$-element
abundance based on three elements is well in agreement  with similar abundances for 
Galactic globular clusters available 
in the literature (see Figure~\ref{fig:clustersep}). Indeed, the mean $\alpha$-element 
abundances for metal--intermediate globulars ($-1.7 \lesssim$ [Fe/H] $\lesssim -1.0$)  
range from $\sim$0.23 (NGC~6205/M13, NGC~6254/M10) to $\sim$0.45 dex (NGC~1904/M79), 
with a dispersion of 0.06 dex\footnote{We neglected  the two extreme 
clusters Rup~106 ([Fe/H]=$-1.36$, [$\alpha$/Fe]$\sim$0) and NGC~6362 ([Fe/H]=$-1.04$, [$\alpha$/Fe]$\sim$0.56). 
Solid evidence based on metallicity distribution and on the absolute 
age suggest that the former one was accreted  \citep{villanova13}.  The 
$\alpha$-elements abundance of the latter is only based on two
stars \citep{gratton87}. However, more recent estimates based on a larger sample 
\citep{massari17} suggest [Fe/H]=$-1.07$ and
[$\alpha$/Fe]=0.37 for NGC~6362, similar to the other metal-intermediate 
globulars.}.

\begin{figure}
\plotone{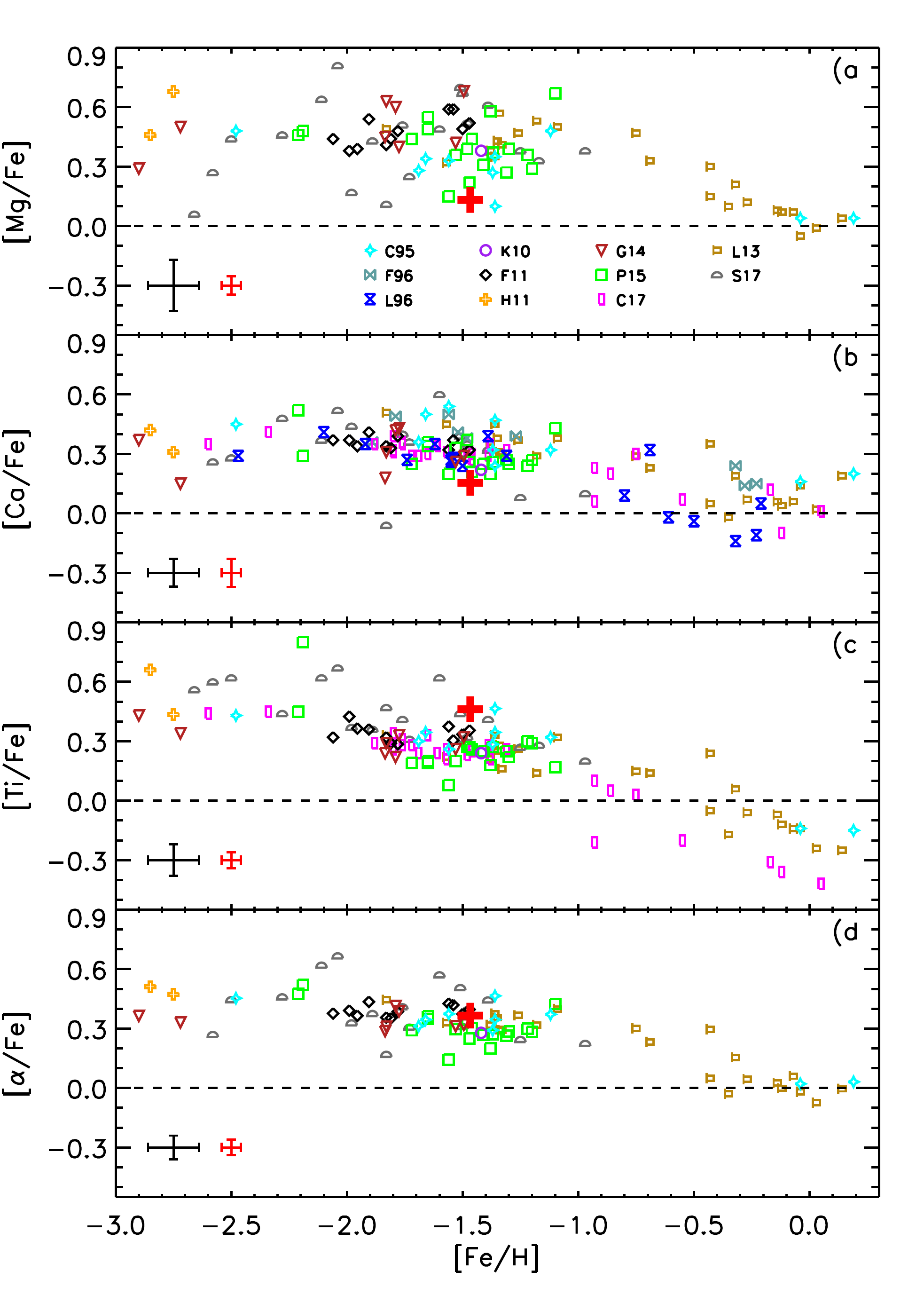}
\caption{$\alpha$-elements vs iron abundances by high-resolution spectroscopy of field Halo RRLs (C95=\citealt{clementini95}; F96=\citealt{fernley96}; L96=\citealt{lambert96}; K10=\citealt{kolenberg10}; F11=\citealt{for11}; H11=\citealt{hansen11}; L13=\citealt{liu13}; G14=\citealt{govea14}; P15=\citealt{pancino15}; C17=\citealt{chadid17}; S17=\citealt{sneden17}; see also Table~\ref{tab:alfalit}). The black error bars on bottom left corner show the mean individual errors. The red cross, with red error bar in the bottom left corner, shows our analysis of NGC~3201. \label{fig:highresrrlsep}}
\end{figure}

To further investigate the possible differences between field and cluster RR Lyrae 
we compared the current $\alpha$-elements abundances with similar ones 
for field RR Lyrae.  Figure~\ref{fig:highresrrlsep} shows the comparison with 
147 field RRLs (96 objects) for which the abundances are based on high resolution spectra. 
They are marked with different symbols and colors and they have been re-scaled to 
the same solar abundances \citep{asplund09}. 
We selected the stars in common among the different spectroscopic samples to 
estimate the standard deviation in iron and $\alpha$-element abundances as 
representative of the individual star errors (see the black error bars plotted in the 
bottom left corners of the figure).
The range in iron abundance covered 
by field RRLs is similar to the globular iron abundances, with a slight overdensity
of stars in the metal-intermediate range ($-1.7 \lesssim$ [Fe/H] $\lesssim -1.0$).
We found that Mg is confirmed to have the highest dispersion  
(0.14 dex). As for globulars, our 
results for NGC~3201 show a limited enhancement of Mg compared with RRLs of similar iron content. 
The other two $\alpha$-elements  
(Ca, Ti) have smaller standard deviations, both $\sim$0.08 dex. The Ca abundance 
is once again slightly under-enhanced, but within the intrinsic dispersion. On the other hand, 
the Ti abundance of RRLs in NGC~3201 seems to be, at fixed iron content, over-enhanced 
when compared with field RRLs. 
The under-- and over-- enhancements of Mg--Ca and Ti mostly balance each other when considering
the [$\alpha$/Fe] ratios, 
so that NGC~3201 is in agreement with field RRLs of similar iron content, as showed in 
the panel d) of Figure~\ref{fig:highresrrlsep}.

\begin{figure}
\plotone{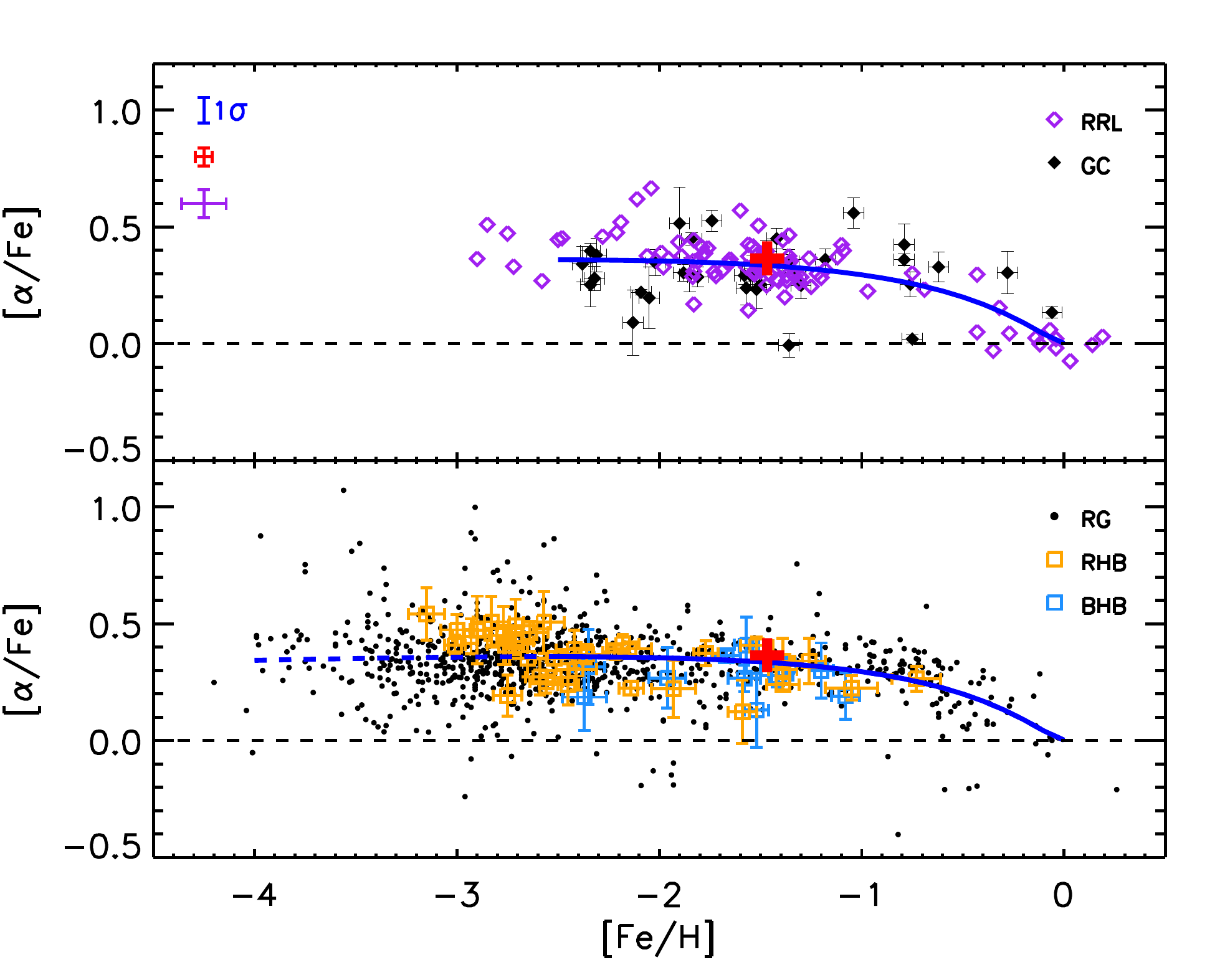}
\caption{Top panel--  $\alpha$-elements (Mg+Ca+Ti) vs iron abundances of Galactic globulars (filled black diamonds) and field Halo RRLs (open purple diamonds, typical error bar is shown on top left corner). The two samples are the same as in panels d) of Figures \ref{fig:clustersep}--\ref{fig:highresrrlsep}. The red cross, with red error bar in the top left corner, shows our analysis of NGC~3201. The solid blue line shows the log-normal fit of the two joint samples, with the 1$\sigma$ dispersion shown by the blue bar in the top left corner. 
Bottom panel-- Comparison of field Halo giants \citep[black dots,][]{frebel10c} and RHB--BHB field stars \citep[orange--blue squares,][]{for10}. The red cross shows our analysis of NGC~3201. The solid blue line shows the log-normal fit of the joint samples of field Halo RRLs and Galactic globulars as in the upper panel. The dashed blue line shows an extrapolation of the fit toward lower iron abundances.  \label{fig:stelledicamposep}}
\end{figure}

To constrain on a more quantitative basis the comparison between field and cluster 
RR Lyrae with field Halo stars we performed an analytical fit of [$\alpha$/Fe] vs [Fe/H]    
for both field RRLs and Galactic globulars. Note that, for these two sub-samples, 
we have solid reasons to believe that their age is similar and  $\ge$10 Gyrs. The fit was 
performed over the range in metallicity $-2.5 <$ [Fe/H] $< 0.0$, since the sampling of the 
more metal-poor regime is limited.  We adopted a log-normal distribution and 
we found

%\begin{equation*}
%[\alpha/Fe] = H \cdot \exp \left[\ln (2)\cdot\frac{\ln^2 \left(1+\frac{2A([Fe/H]-[Fe/H]_0)}{\sigma}\right)}{A^2}\right]
%\end{equation*}

\begin{equation}
[\alpha/Fe] = H \exp \left[\ln (2) \frac{X^2}{A^2}\right]
\end{equation}

\noindent with

\begin{equation}
X = \ln \left[1+\frac{2A([\text{Fe/H}]-[\text{Fe/H}]_0)}{\sigma} \right]
\end{equation}

\noindent where $H$(scale height)$=0.359438$, 
representing the [$\alpha$/Fe] abundance for very metal poor ([Fe/H]$\lesssim$$-$2) stars,
$A$(asymmetry)=$-1.74038$, [Fe/H]$_0$=$-2.56024$ and $\sigma$=8.98116.
The top panel of Figure~\ref{fig:stelledicamposep} shows the analytical fit (blue line), 
together with the quoted sub-samples. The bottom panel of the same figure shows the 
comparison of the quoted analytical fit, extrapolated down to [Fe/H]=$-$4 (dashed blue line), 
with the mean [$\alpha$/Fe] abundance for NGC~3201
(red cross), the kinematically selected field Halo giants (black dots) collected by 
\cite{frebel10c} and field Halo blue (blue squares, BHB) and red (orange squares, RHB) 
HB stars collected by \cite{for10}. Interestingly enough, the different samples 
do agree within 1$\sigma$, thus suggesting a very similar chemical enrichment history
even though they cover different ranges in iron content and in Galactocentric distances.  This finding further 
supports a common old (t $\ge$ 10 Gyr) age for field RG and HB stars. 

%_______________________________________________________________________________
\subsection{Iron peak elements: Sc, Cr, Ni, Zn} \label{sec:ironpeak}
%_______________________________________________________________________________

Iron peak elements are defined as those with Z=21--30.
We measured spectral lines of four iron peak elements: 
Sc, Cr, Ni, Zn (see Table~\ref{tab:ironpeak} for details).
Chromium is the most represented element in our sample, since it was measured 
in eleven out of twelve stars, with up to ten lines in the RHB star. 
The other three elements (Sc, Ni, Zn) were observed in only a half of the 
current sample with a limited number of lines. 
Our derived mean abundances are 
$\langle$[Sc/Fe]$\rangle$=$0.14\pm0.06$, 
$\langle$[Cr/Fe]$\rangle$=$0.12\pm0.05$,
$\langle$[Ni/Fe]$\rangle$=$-0.02\pm0.04$ and 
$\langle$[Zn/Fe]$\rangle$=$0.22\pm0.06$.

\begin{figure}
\plotone{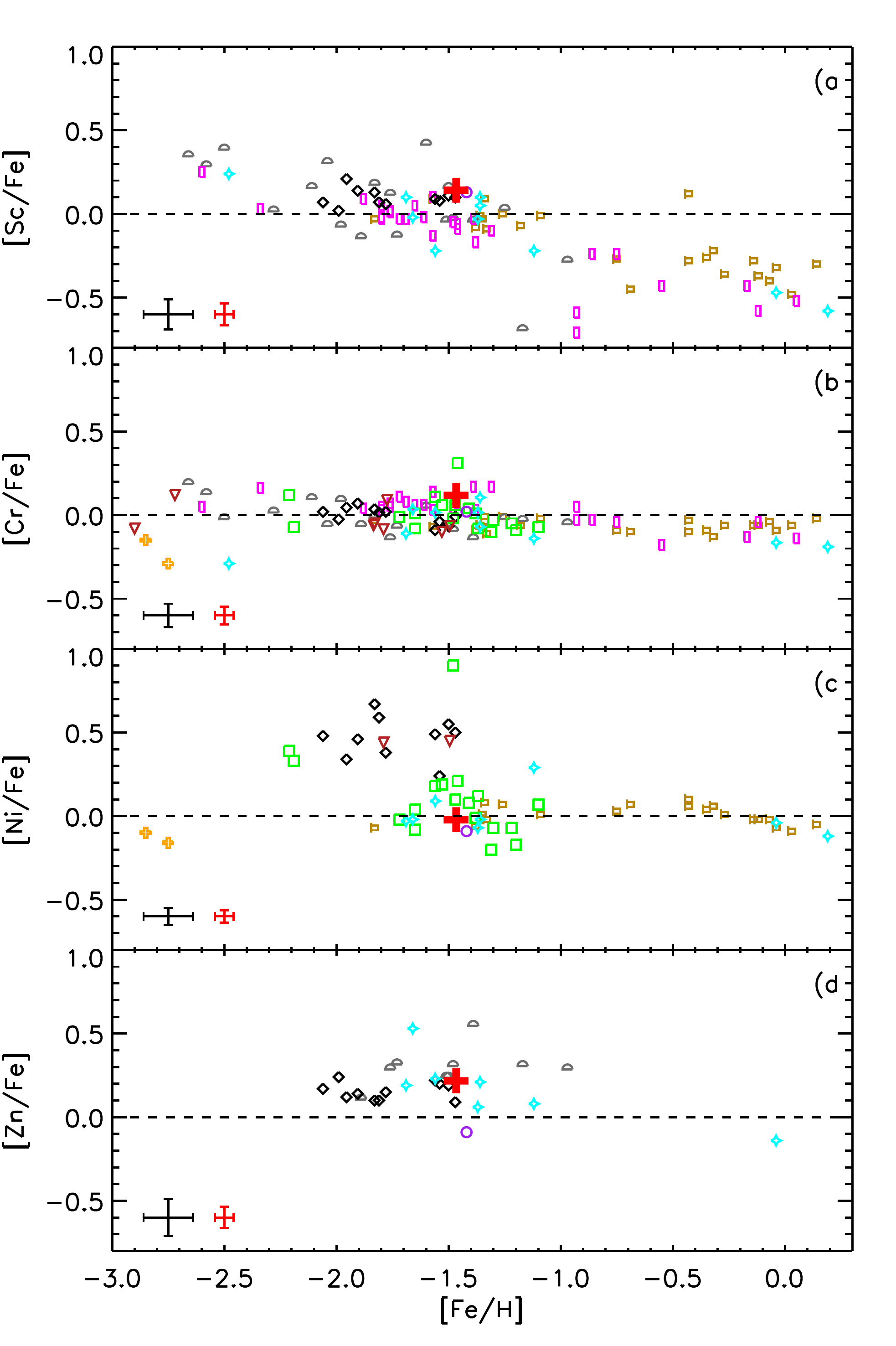}
\caption{Iron-peak elements vs iron abundance by high-resolution spectroscopy of field Halo RRLs 
(same symbols as in Figure~\ref{fig:highresrrlsep};
see also Table~\ref{tab:heavylit}). The black error bars on bottom left corner show the mean individual errors. The red cross, with red error bar in the bottom left corner, shows our analysis of NGC~3201.  \label{fig:highresheavy}}
\end{figure}

Figure~\ref{fig:highresheavy} shows these abundances
and those for field Halo RRLs.
The Fe-group abundance ratios in NGC~3201 generally agree with those of 
the other RRLs.
Nickel abundances deserve some comment.  
Our [Ni/Fe] values agree with most of the previous studies, but the 
abundances by \cite{for11} and by \cite{govea14} appear to be enhanced by 
$\sim$0.5 dex.
\cite{for11} noted that the phase to phase scatter in Ni abundances, 
for which they only measured a couple of lines, should be treated with caution.  
This scatter can be seen in their Figures~23 and 26.  
\cite{govea14} measured only one Ni line and in a single phase for each 
star, so their estimates have larger uncertainties.  
We conclude that our Fe-group abundances are in agreement with other 
stellar samples in this metallicity regime.

\begin{deluxetable*}{crlcrlcrlcrlc}
\tablecaption{Iron peak elements abundances and number of lines. \label{tab:ironpeak}}
\tablewidth{0pt} 
\tablehead{
\colhead{ID} &
\twocolhead{[Sc/Fe]} &
\colhead{n} &
\twocolhead{[Cr/Fe]} &
\colhead{n} &
\twocolhead{[Ni/Fe]} &
\colhead{n} &
\twocolhead{[Zn/Fe]} &
\colhead{n}
}
\startdata
V3       &    \nodata &                & \nodata &     0.06  & $\pm$    0.14  &      3  &    -0.04  & $\pm$    0.11  &      2  &     0.42  &                &      1   \\
V6       &    \nodata &                & \nodata &     0.47  &                &      1  &   \nodata &                & \nodata &   \nodata &                & \nodata  \\
V14      &    \nodata &                & \nodata &     0.27  &                &      1  &   \nodata &                & \nodata &   \nodata &                & \nodata  \\
V26      &      0.20  & $\pm$    0.02  &      2  &     0.19  & $\pm$    0.02  &      5  &    -0.06  & $\pm$    0.12  &      2  &     0.35  & $\pm$    0.08  &      2   \\
V37      &      0.12  & $\pm$    0.00  &      2  &     0.20  &                &      1  &   \nodata &                & \nodata &   \nodata &                & \nodata  \\
V38      &     -0.04  & $\pm$    0.16  &      2  &    -0.05  & $\pm$    0.11  &      3  &    -0.10  & $\pm$    0.07  &      2  &     0.19  &                &      1   \\
V41      &      0.15  & $\pm$    0.11  &      3  &     0.00  & $\pm$    0.02  &      5  &    -0.11  & $\pm$    0.09  &      2  &     0.23  & $\pm$    0.06  &      2   \\
V47      &      0.05  &                &      1  &    -0.15  & $\pm$    0.10  &      2  &     0.17  &                &      1  &   \nodata &                & \nodata  \\
V57      &      0.21  & $\pm$    0.10  &      2  &     0.02  & $\pm$    0.02  &      2  &     0.04  &                &      1  &     0.13  &                &      1   \\
V73      &     -0.17  &                &      1  &   \nodata &                & \nodata &   \nodata &                & \nodata &   \nodata &                & \nodata  \\
V83      &      0.53  &                &      1  &     0.25  & $\pm$    0.13  &      3  &   \nodata &                & \nodata &   \nodata &                & \nodata  \\
94180    &      0.23  & $\pm$    0.04  &      2  &     0.04  & $\pm$    0.05  &      9  &    -0.04  & $\pm$    0.04  &      5  &    -0.02  & $\pm$    0.12  &      2   \\
\hline
NGC 3201 &      0.14  & $\pm$    0.06  &         &     0.12  & $\pm$    0.05  &         &    -0.02  & $\pm$    0.04  &         &     0.22  & $\pm$    0.06  &          \\
\enddata
\end{deluxetable*}

\begin{figure}
\plotone{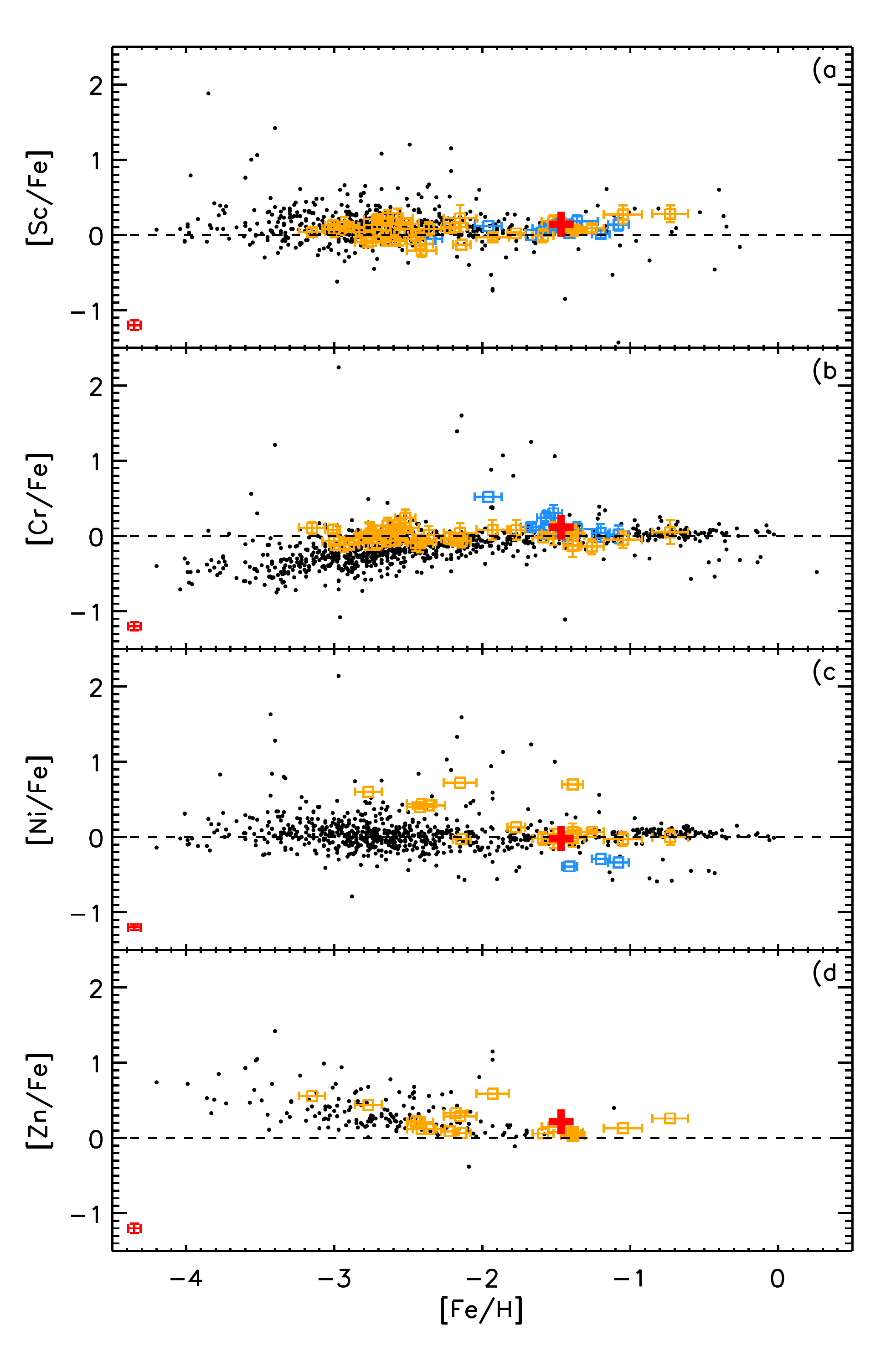}
\caption{Iron-peak elements vs iron abundance of field Halo giants and RHB--BHB field stars (same symbols as in Figure~\ref{fig:stelledicamposep}). The red cross, with red error bar in the bottom left corner, shows our analysis of NGC~3201.  \label{fig:stelledicampo_heavy}}
\end{figure}

In Figure~\ref{fig:stelledicampo_heavy} we show our 
Fe-group abundances and those of field Halo giants collected by 
\cite{frebel10c} and Halo RHB--BHB stars by \cite{for10}. 
All the elements in NGC~3201 are in good agreement with other stars
in the intermediate metallicity regime.
This suggests once again that field Halo RGs, field HBs (RHB, RRL, BHB) 
and globular clusters share similar chemical enrichments.

%_______________________________________________________________________________
\subsection{s-process element: Yttrium} \label{sec:y}
%_______________________________________________________________________________

The slow neutron-capture process ($s$-process), 
in which timescales for capture of free neutrons
are longer than timescales of $\beta$-decays, has dominated the production 
of yttrium in solar-system material. 
However, the $s$-process fractional dominance over the $r$-process (rapid 
neutron-capture) that produced solar Y is still an open question.
\cite{simmerer04} estimated a fraction of $\sim$72\%, whereas 
\cite{arlandini99} estimated much higher values of $\sim$92--100\%.

\begin{figure}
\plotone{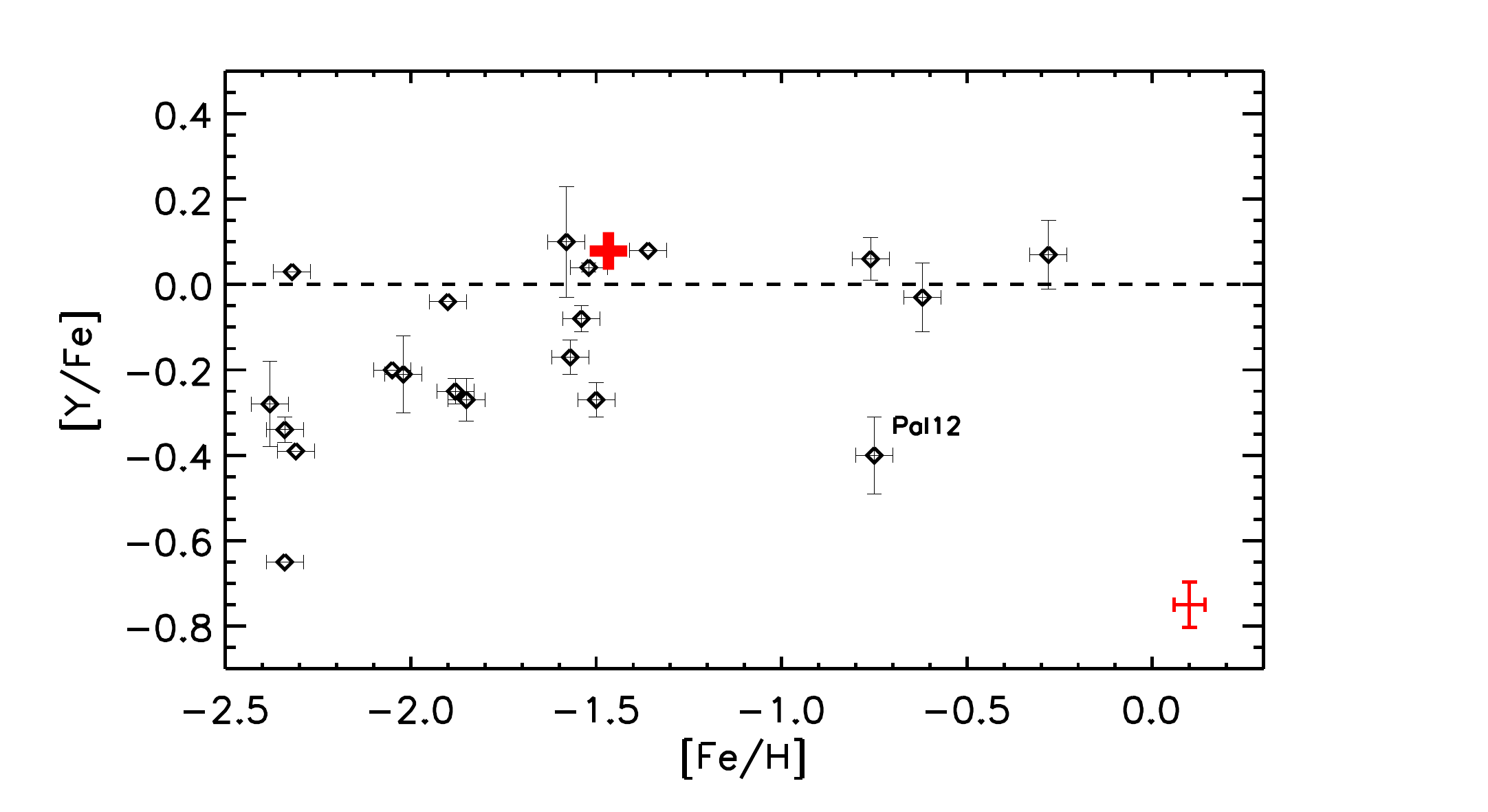}
\caption{Yttrium vs iron abundance of globular clusters \citep{pritzl05b}. The red cross, with red error bar in the bottom right corner, shows our analysis of NGC~3201.  \label{fig:ycluster}}
\end{figure}

\begin{figure}
\plotone{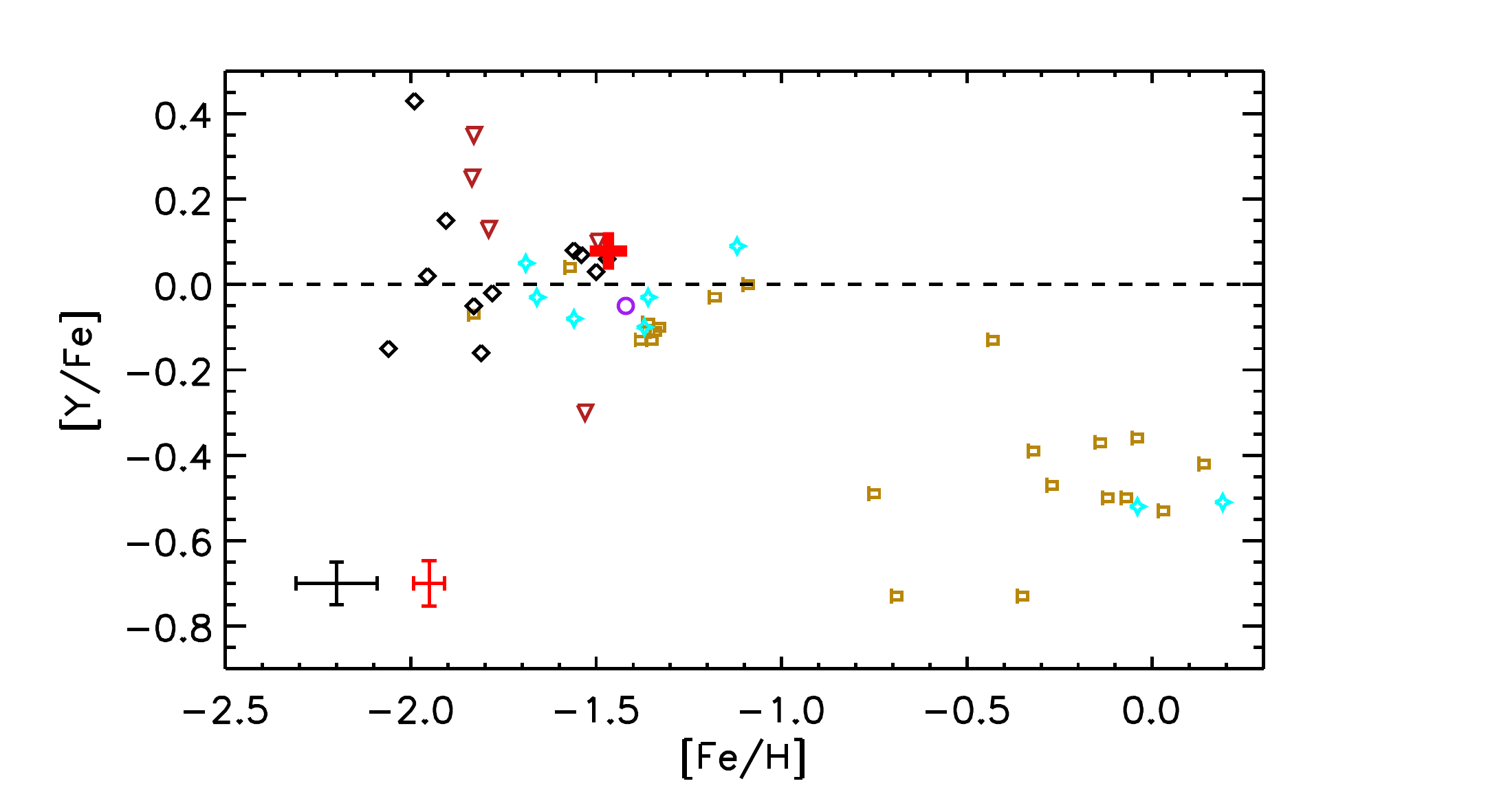}
\caption{Yttrium vs iron abundance by high-resolution spectroscopy of field Halo RRLs 
(symbols as in Figure~\ref{fig:highresrrlsep};
see also Table~\ref{tab:heavylit}). The black error bar on bottom left corner shows the mean individual error. The red cross, with red error bar in the bottom left corner, shows our analysis of NGC~3201.  \label{fig:highresY}}
\end{figure}

Within the M2FS spectral range, we identified five potentially
useful Y~II lines, and measured up to four EWs in nine out of our twelve stars
(see Table~\ref{tab:ittrio}). 
From these we estimated an average abundance for NGC~3201 of
$\langle$[Y/Fe]$\rangle$=0.08$\pm$0.05.
Figure~\ref{fig:ycluster} shows the comparison of this result and average Y
abundances of other Galactic globulars \citep{pritzl05b}.
The cluster-to-cluster scatter in [Y/Fe] appears to be large, but this
could simply reflect study-to-study differences in the \cite{pritzl05b}
compilation.
Relatively  metal rich clusters show a small dispersion and an average 
[Y/Fe] abundance close to solar, with the exception of the outer Halo cluster 
Pal~12. 
The peculiarity of this cluster is not surprising, since there are 
photometric and spectroscopic reasons to believe that Pal~12 is an accreted 
cluster \citep[][and references therein]{musella18}. 
Clusters with [Fe/H]~$\lesssim$~$-$1.8 often exhibit sub-solar [Y/Fe] 
values, but the scatter is still large.
Field Halo RRLs yield similar [Y/Fe] results in
the metal-intermediate ($-1.7 \lesssim$ [Fe/H] $\lesssim -1$) regime
(Figure~\ref{fig:highresY}).
Comparison of the data for globular clusters and field 
stars shows mostly that 
a future study is needed to bring coherence to [Y/Fe] abundance trends with 
metallicity.

\begin{deluxetable}{crlc}
\tablecaption{Ittrium abundances and number of lines. \label{tab:ittrio}}
\tablewidth{0pt} 
\tablehead{
\colhead{ID} &
\twocolhead{[Y/Fe]} &
\colhead{n} 
}
\startdata
V3       &  -0.01  & $\pm$   0.15  &      2   \\  
V6       & \nodata &               & \nodata  \\  
V14      &   0.28  & $\pm$   0.10  &      2   \\  
V26      &   0.24  & $\pm$   0.01  &      2   \\  
V37      & \nodata &               & \nodata  \\  
V38      &  -0.14  & $\pm$   0.09  &      2   \\  
V41      &   0.04  & $\pm$   0.06  &      4   \\  
V47      &  -0.17  & $\pm$   0.07  &      2   \\  
V57      &   0.15  & $\pm$   0.11  &      2   \\  
V73      &   0.15  & $\pm$   0.04  &      2   \\  
V83      & \nodata &               & \nodata  \\  
94180    &   0.17  & $\pm$   0.14  &      3   \\  
\hline                                       
NGC 3201 &   0.08  & $\pm$   0.05  &          \\   
\enddata
\end{deluxetable}

\begin{figure}
\plotone{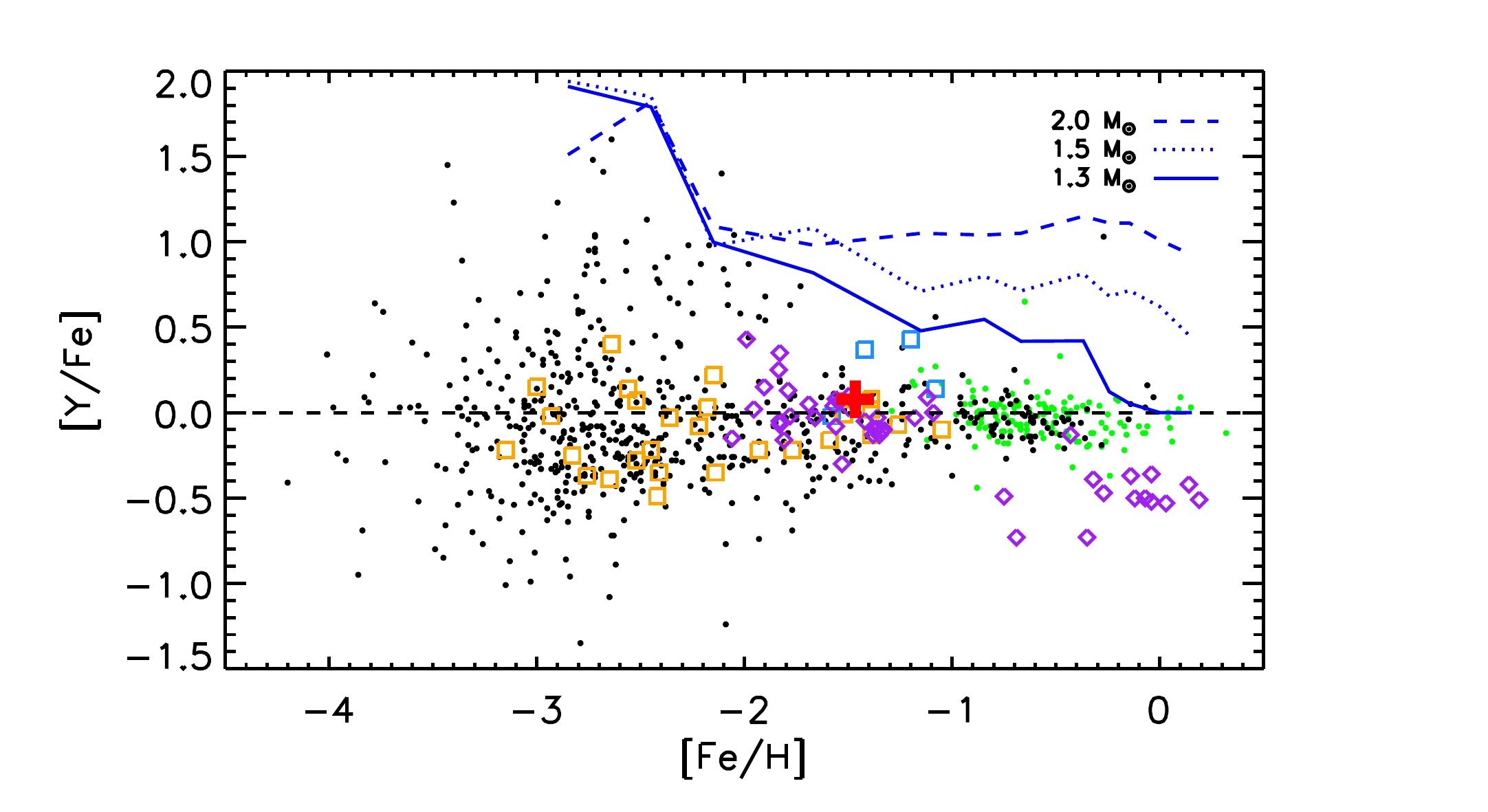}
\caption{Yttrium vs iron abundances of field Halo giants \citep[black dots,][]{frebel10c}, RHB--BHB field stars \citep[orange--blue squares,][]{for10}, Thin--Thick Disk stars \citep[green dots,][]{reddy06} and high-resolution spectroscopic RRLs (purple diamonds, same sample as in Figure~\ref{fig:highresY}). The red cross shows our analysis of NGC~3201. The blue lines show theoretical predictions for different solar masses, based on the FRUITY database \citep{cristallo11,cristallo15}.} \label{fig:halo-disk}
\end{figure}

Of special interest is the behavior of [Y/Fe]
in the metal-rich ([Fe/H] $\gtrsim -1.0$) regime.
Field RRLs shows a severe depletion of Y ([Y/Fe]=$-0.48\pm0.15$;
Figure~\ref{fig:highresY}), but ones in clusters do not 
(Figure~\ref{fig:ycluster}).
The metal-rich RRLs mainly come from the dataset collected by \cite{liu13}, 
with the exception of two stars provided by \cite{clementini95}. 
The two stars by \citeauthor{clementini95} are also in the dataset by 
\citeauthor{liu13}, with very similar derived abundances,
excluding the possibility of systematics in one of 
the two samples. 
Considerations about their radial velocities suggest 
that these RRLs are candidate members of the Galactic disk, not the Halo 
(\citealt{liu13}). 
To further investigate RRLs in the metal-rich regime, we compared
their abundances with Halo stars. In Figure~\ref{fig:halo-disk} we plot
[Y/Fe] ratios versus metallicity for various samples of stars.
In the metal-poor domain ([Fe/H] $\lesssim -1.0$), the field RRLs show  
agreement, on average, with other field Halo stars, namely RG and HB stars. 
On the contrary, the metal rich  ([Fe/H]$\gtrsim-1.0$) tail of RRLs is 
clearly depleted in Y when compared with slightly more metal-poor 
([Fe/H]$\sim-1.0$) field Halo stars and with Disk dwarfs
of similar metallicities \citep{reddy06}. 
These results, coupled with the radial velocity considerations by \cite{liu13},
suggest that they may be candidate Bulge members.

In Figure~\ref{fig:halo-disk} we also include [Y/Fe]
predicted trends from evolutionary prescriptions based on
Asymptotic Giant Branch models available on the FRUITY\footnote{
\url{http://fruity.oa-teramo.inaf.it/}} 
database \citep{cristallo11,cristallo15}.
Blue lines in the figure show predicted values for three different 
stellar masses (see labelled values). 
Unfortunately, for masses $\leq$1.1 M$_\odot$,
as the RRLs have, the accounted models
are not available because they do not experience the
third dredge up and, therefore, do not show a significant 
chemical enrichment of the environment.
However, the theoretical predictions for the less 
massive available stars show a rapid decrease in Y abundance
for [Fe/H]$\gtrsim-0.45$, suggesting that we are moving in the right direction.

%_______________________________________________________________________________
\section{Conclusion and final remarks} \label{sec:fine}
%_______________________________________________________________________________

We performed the first high resolution, 
high SNR, large spectroscopic investigation of RRLs in NGC~3201.
Our independent analysis confirmed many previous results 
on the cluster using non-variable stars.
Our derived average metallicity of the cluster, 
$\langle$[Fe/H]$\rangle$=$-1.47\pm0.04$, is in general accord with recent
studies, and we confirm that NGC~3201 is a homogeneous, mono-metallic cluster.

A limited dispersion was also observed for three different groups
of elements: the light $\alpha$-elements Mg--Ca--Ti, the iron peak elements
Sc--Cr--Ni--Zn and the s-process element Y.
The $\alpha$-elements were found to be enhanced with respect to the Sun,
as expected for old (t$\ge$10 Gyr) stellar structures, with abundances
comparable to other known globular clusters and field stars of similar 
metallicity.
In particular, the agreement was found not only with HB stars (RHB, RRL, BHB) 
as the ones in our sample, but in general with variable and non-variable 
field Halo stars.
We found $\langle$[Mg/Fe]$\rangle$=0.13$\pm$0.05, $\langle$[Ca/Fe]$\rangle$=0.15$\pm$0.07 and $\langle$[Ti/Fe]$\rangle$=0.46$\pm$0.04.
The same homogeneity was observed for the iron peak and s-process elements, 
whose abundance ratios are close to solar and similar to other 
metal-intermediate ($-1.7 \lesssim$ [Fe/H] $\lesssim -1.0$) globulars and 
field Halo stars.
These results suggests similar enrichment histories for all the analysed
Halo components with similar metallicity to NGC~3201.

The cluster radial velocity was estimated as 494$\pm$2$\pm$8 km~s$^{-1}$,
where the two errors are the error on the mean and the standard deviation, 
respectively.
The use of a template to obtain a radial velocity curve from
a single epoch velocity measurement of a RRL star is a very promising approach.
However, to obtain more precise results it is necessary a good
photometric dataset, almost coeval to the spectroscopic data.
The goodness of this approach is further supported by the homogeneity
of the average template velocities of the RRLs in our sample and the
non-variable star instantaneous velocity.

The results obtained with this work on NGC~3201 strongly
supports the capabilities of M2FS at Magellan as a high quality
instrument for abundance investigations, even in crowded fields such
as a globular cluster.
This work on NGC~3201 is opening the path to a
forthcoming analysis on the more complex Globular
cluster $\omega$~Cen, which is known to have an intrinsic metallicity spread, for which we already collected
M2FS spectra of $\sim$140 RRLs.

\acknowledgments

C.S. was partially supported by NSF grant AST-1616040, and 
by the Rex G. Baker, Jr. Endowment at the University of Texas.
C.S. also thanks the Dipartimento di Fisica -- Universit{\`a} di Roma Tor Vergata 
for a Visiting Scholar grant and INAF -- Osservatorio Astronomico di Roma for his support during his stay.
M.M. was partially supported by NSF grant AST-1714534.
It is a real pleasure to thank P.B. Stetson for sending us in advance of publication optical photometry for the variables included in this investigation. 
We also thank Ivan Ferraro and Giacinto Iannicola 
of the INAF -- Osservatorio Astronomico di Roma for the help 
in computing analytical relations for $\alpha$-element abundances in the Halo.
Moreover, many thanks to Sergio Cristallo for the fruitful discussion about r-
and s- process elements and for sending us the models about Yttrium
abundances computed with the FRUITY database.
Finally, we thank the reviewer for the precise notes and the useful suggestions to improve our work.

This work has made use of the VALD database (\url{http://vald.astro.univie.ac.at/~vald3/php/vald.php}), 
operated at Uppsala University, the Institute of Astronomy RAS in Moscow, and the University of Vienna.

\appendix
%\section{Appendix}

\begin{deluxetable*}{lccccc}
\tablecaption{Iron and $\alpha$-elements abundances of high-resolution RRLs \label{tab:alfalit}}
\tablewidth{0pt}
\tablehead{
\colhead{ID} &	
\colhead{[Fe/H]} &
\colhead{[Mg/Fe]} &
\colhead{[Ca/Fe]} &
\colhead{[Ti/Fe]} &
\colhead{Reference}
}
\startdata
AA Aql               &       -0.32     &   0.21     &  0.19     &    0.06     &   L13   \\
AE Dra               &       -1.46     &   0.44     &  0.21     &    0.26     &   P15   \\
AN Ser               &       0.05      &   \nodata  &  0.01     &    -0.42    &   C17   \\
AO Peg               &       -1.26     &   0.47     &  0.37     &    0.26     &   L13   \\
AR Per               &       -0.23     &   \nodata  &  -0.11    &    \nodata  &   L96   \\
AR Per               &       -0.28     &   \nodata  &  0.14     &    \nodata  &   F96   \\
AR Per               &       -0.29     &   \nodata  &  \nodata  &   \nodata   &   A18   \\
AS Vir               &       -1.57     &   \nodata  &  0.31     &    0.21     &   C17   \\
\enddata
\tablenotetext{}{References: C95=\cite{clementini95}; L96=\cite{lambert96}; K10=\cite{kolenberg10}; F11=\cite{for11}; H11=\cite{hansen11}; L13=\cite{liu13}; G14=\cite{govea14}; P15=\cite{pancino15}; C17=\cite{chadid17}; S17=\cite{sneden17}. All scaled to \cite{asplund09}.}
\tablenotetext{}{NOTE---Table \ref{tab:alfalit} is published in its entirety in the machine readable format. A portion is shown here for guidance regarding its form and content.}
\end{deluxetable*}

\begin{deluxetable*}{lcccccc}
\tablecaption{Iron peak and s-process elements abundances of high-resolution RRLs \label{tab:heavylit}}
\tablewidth{0pt}
\tablehead{
\colhead{ID} &	
\colhead{[Sc/Fe]} &
\colhead{[Cr/Fe]} &
\colhead{[Ni/Fe]} &
\colhead{[Zn/Fe]} &
\colhead{[Y/Fe]} &
\colhead{Reference}
}
\startdata
AA Aql                 &   -0.22      &   -0.13     &  0.06       &  \nodata  &  -0.39   &   L13   \\
AE Dra                 &   \nodata    &   0.31      &  0.21       &  \nodata  &  \nodata &   P15   \\
AN Ser                 &   -0.52      &   -0.14     &  \nodata    &  \nodata  &  \nodata &   C17   \\
AO Peg                 &   0.00       &   -0.01     &  0.07       &  \nodata  &  \nodata &   L13   \\
AS Vir                 &   -0.13      &   0.14      &  \nodata    &  \nodata  &  \nodata &   C17   \\
AS Vir                 &   0.06       &   0.02      &  0.38       &  0.15     &  -0.02   &   F11   \\
ASAS J081933-2358.2    &   \nodata    &   -0.08     &  \nodata    &  \nodata  &  \nodata &   G14   \\
ASAS J085254-0300.3    &   0.17       &   -0.05     &  \nodata    &  0.25     &  \nodata &   S17   \\
\enddata
\tablenotetext{}{References: C95=\cite{clementini95}; K10=\cite{kolenberg10}; F11=\cite{for11}; H11=\cite{hansen11}; L13=\cite{liu13}; G14=\cite{govea14}; P15=\cite{pancino15}; C17=\cite{chadid17}; S17=\cite{sneden17}. All scaled to \cite{asplund09}.}
\tablenotetext{}{NOTE---Table \ref{tab:heavylit} is published in its entirety in the machine readable format. A portion is shown here for guidance regarding its form and content.}
\end{deluxetable*}

%\begin{thebibliography}{}
\bibliographystyle{aasjournal}
\bibliography{ms}
%\end{thebibliography}

\end{document}